\documentclass[%
 reprint,
prb,
]{revtex4-2}

\usepackage{amsmath,amssymb}
\usepackage{graphicx}
\usepackage{dcolumn}
\usepackage{bm}
\usepackage{xcolor}
\usepackage{siunitx}
\usepackage{hyperref}

\begin{document}

\preprint{APS/123-QED}

\title{Valley Order in Moir\'e Topological Insulators}
\author{Bo Zou*}
\thanks{These authors contributed equally to this work.}
\affiliation{Department of Physics, University of Texas at Austin, Austin, TX 78712}
\author{Anzhuoer Li*}
\thanks{These authors contributed equally to this work.}
\affiliation{Department of Physics, University of Texas at Austin, Austin, TX 78712}

\author{A. H. MacDonald}

\affiliation{Department of Physics, University of Texas at Austin, Austin, TX 78712}

\date{\today}

\begin{abstract}
Moir\'e materials with opposite non-zero miniband 
Chern numbers in time-reversal-partner valleys
are two-dimensional topological insulators at band filling $\nu=2$.  
We explore the possibility that in this class of moir'e materials intervalley coherence 
can sometimes be present in interaction induced insulators at band 
filling $\nu=1$ , using Landau levels with opposite signs of the magnetic 
field as a convenient generic model.  
In the absence of intravalley interactions the mean-field ground 
state at filling factor $\nu=1$ is a gapless intervalley coherent state
that maps under a particle-hole transformation of one valley to a strong-field 
superconducting vortex-lattice state that has been studied previously.  
When the ratio $\lambda$ of intravalley to intervalley interactions is increased, 
gapped states appear, one with broken time-reversal symmetry and a quantized Hall effect but no valley polarization and one with broken parity symmetry and zero Hall conductivity. 
We discuss the possibility that the latter state could be related to the fractional quantum spin Hall effect recently observed at an odd filling factor in a moir'e topological insulator and comment on related systems in which correlations between electrons in bands with opposite Chern numbers might play a key role.  
\end{abstract}

\maketitle


\section{Introduction}
The band edges of both graphene and group VI transition-metal-dichalcogenide (TMD) 
two-dimensional semiconductors are located at the $K$ and $K'$ corners of 
their triangular lattice Brillouin zone.  
A key property of these momenta is that they are not time-reversal
invariant, which allows the valley projected moir\'e minibands formed when two or more
layers are overlaid with a small twist or difference in lattice constant
to have non-zero Chern numbers.
Because of time-reversal symmetry, opposite valleys must have opposite Chern numbers.
It follows that whenever the non-zero Chern number case is realized, 
the twisted bilayer is a two-dimensional topological insulator \cite{kane2005quantum} when 
the band filling factor $\nu=2$, {i.e.} when 
both bands are occupied.  This property has recently been confirmed experimentally \cite{kang2024double}.   In the case of twisted MoTe$_2$ and WSe$_2$ K-valley homobilayers,
which have received a lot of recent attention \cite{wang2020correlated, cai2023signatures, zeng2023thermodynamic, xu2023observation, park2023observation, foutty2024mapping, ji2024local, kang2024evidence, redekop2024direct, gao2025probing, guo2025superconductivity, xia2025superconductivity, xu2025interplay},
the non-zero valley Chern numbers arise from the position and momentum dependence of layer 
pseudospins \cite{wu2019topological}
 \footnote{Because spin and valley are locked in these strongly spin-orbit coupled 
 semiconductors, we refer to spin and valley interchangeably.  Following the common convention in moir\'e materials we define band filling factor $\nu$ as the number of electrons per moir\'e period so that spin/valley degenerate bands are full at even integer values of $\nu$ and half-filled at odd integer values.}. 
Graphene multilayers that have a twist or are aligned to hexagonal 
boron nitride have broadly similar properties \cite{cao2018unconventional, han2024large, liu2020tunable, lu2024fractional, serlin2020intrinsic}.
In this article, we will refer to K/K'-valley 
moir\'e materials with non-zero valley-projected Chern numbers as moir\'e topological insulators.

At odd integer filling factors moir\'e topological insulators often 
exhibit quantized anomalous Hall 
effects \cite{chang2023colloquium,sharpe2021evidence,tao2022valley, kang2024evidence} 
that are naturally explained as interaction-induced spontaneously valley-polarized 
insulators.  (In the absence of interactions all
$\nu=1$ states are metallic and have large Fermi surfaces.)
Here we address the competition of these states 
with competing valley-coherent \cite{nuckolls2023quantum} states, which can be 
gapped or have isolated band-touching Dirac points.
Some similar issues have been addressed recently in Refs.~\cite{kwan2025textured,wang2025chern}.
With the goal of capturing behaviors that are generic to moir\'e topological 
insulators that have narrow bands with
nearly ideal quantum geometry, like MoTe$_2$ and WSe$_2$
\cite{devakul2021magic,morales-duran2023pressureenhanced}, 
we replace the two degenerate valley-projected Chern bands by Landau levels with opposite 
signs of magnetic field, which have the required band geometry.
A similar approach was taken previously in Ref.~\cite{bultinck2020mechanism}.
We find that gapped states with quantized anomalous Hall effects 
can be induced not only by valley polarization, but also by valley coherence that breaks 
time-reversal symmetry.  Inversion-symmetry-breaking gapped states also appear that do not 
break time-reversal symmetry and therefore have zero quantum anomalous Hall conductivity.
The relative energies of competing many-body ground states are 
controlled primarily by the ratio $\lambda$ of the intravalley interaction strength to the 
intervalley interaction strength.  In our calculations, valley coherent states appear only when intravalley interactions are weaker than intervalley interactions.

Recent measurements \cite{kang2024evidence} have yielded evidence that 
the fractional quantum spin Hall effect (FQSHE), a fractionalized version of the 
ordinary spin Hall effect \cite{kane2005quantum}, occurs in twisted MoTe$_2$ (tMoTe$_2$)
at small twist angles $\theta$ and odd-integer band filling factors $\nu$ 
\footnote{There are also experimental results \cite{li2025universal} opposing this conclusion.}.
In our theory study, we will focus on $\nu=1$, assuming that only one partially filled band has active degrees of freedom.
The FQSHE observations were not anticipated theoretically, but a large number of interesting potential explanations \cite{abouelkomsan2025nonabelian, ahn2024nonabelian, chen2025robust, chou2024composite, crepel2024spinon, jian2025minimal, may-mann2025theory, reddy2024nonabelian, sodemannvilladiego2024halperin, wang2025higher, xu2025multiple, zhang2024nonabelian, kwan2024could} have been advanced.  Many of these rely on the appearance of 
exotic underlying electronic ground states; the simplest possible explanation, for example, is that the observation is literally evidence for a many-body state that is the direct product of separate $\nu=1/2$ fractional quantum Hall states with 
opposite signs of Hall conductivity for opposite valleys -
a state that does not have inter-valley correlations and is therefore favored by weak 
intervalley interactions.  Theories of the FQSHE must explain
i) the appearance of a gapped state with no anomalous Hall effect (AHE) at an odd integer value of $\nu$, 
ii) dissipationless edge transport with conductivity $e^2/2h$ per edge, and 
iii) the emergence of a small non-quantized Hall effect in the presence of a perpendicular 
magnetic field that has opposite signs on opposite sides of integer $\nu$.
In our study, we conclude that time-reversal invariant inter-valley coherent states cannot on their own explain the FQSHE observations, although they satisfy some necessary conditions.

This paper is organized as follows.  Section II details
the calculations we have performed and Section III describes the results we have 
obtained. In Section IV we discuss some implications of our results as they relate to the fractional quantum spin Hall effect observations \cite{kang2024evidence}
and to bilayer systems with opposite valley polarizations in the two layers.
Finally in Section V we present our conclusions.  Some technical details are presented 
in Appendices.  

\section{Mean-Field Theory of Moiré Topological Insulators}

We are interested in properties that are generic to
flat-band moiré topological insulators, defined as systems with locked
spin/valley degrees of freedom and opposite Chern numbers for flat bands with 
opposite spin.  A system with two perfectly flat Landau levels that experience 
opposite signs of magnetic field for opposite spins, provides a typical example 
of such a system.  In MoTe$_2$ homobilayer moir\'es, which map under an adiabatic approximation \cite{morales2024magic,shi2024adiabatic} to Aharonov-Casher bands,
the wavefunctions of the flat bands are especially similar to Landau-level bands since
they accurately approximate the ideal quantum geometry \cite{wang2021exact, wang2023origin}
of Landau levels.  
The similarity of Landau-level wavefunctions and MoTe$_2$ homobilayer moir\'e
wavefunctions has been verified by DFT calculations \cite{zhang2024polarization, wang2025higher} for 2.1$^\circ$ tMoTe$_2$.
We take the view that results calculated with the Landau level model we employ
are representative of typical moiré topological insulators, 
and that strong correlation physics is more likely to be
manifested in TMD homobilayer moir\'es when the wavefunctions are 
Landau-level-like, justifying our choice of a model system.  

In the representation of Landau-level quasi-Bloch states\cite{brown1964bloch,zak1964magnetic,haldane1985many,haldane2018origin, apxquasibloch}, the interaction Hamiltonian
\begin{equation} \label{eq:coulomb interaction}
\begin{aligned}
    H_{\text{int}} =& \frac{1}{2A} \sum_{\textbf{k},\textbf{p}}^\text{BZ} \sum_{\textbf{q}} \sum_{s_1, s_2} V(\textbf{q})     \left< s_1\textbf{k}+\textbf{q} \middle| e^{i\textbf{q}\cdot\textbf{r}} \middle| s_1 \textbf{k} \right> \\
    &\left< s_2\textbf{p}-\textbf{q} \middle| e^{-i\textbf{q}\cdot\textbf{r}} \middle| s_2 \textbf{p} \right> 
    c^\dagger_{s_2\textbf{p}-\textbf{q}} c^\dagger_{s_1\textbf{k}+\textbf{q}} c_{s_1\textbf{k}} c_{s_2\textbf{p}}.
\end{aligned}
\end{equation}
Here $s_i$ labels valley and we have projected the interacting problem onto a 
Hilbert space with one band for each valley/spin.  In Eq.~\ref{eq:coulomb interaction}, $V(\textbf{q}) = (2\pi e^2/\epsilon|\textbf{q}|)\cdot\tanh{qD}$ is the Fourier transform of a Coulomb interaction screened by top and bottom gates removed by distance $D$, $\epsilon\approx 5$ is the dielectric constant of MoTe$_2$, and $A$ is the sample area.
For explicit calculations, we take $D$=20 nm, so that
gate screening has a modest quantitative effect on our results. 
For the Landau levels, we must choose the lattice unit cell so that its area $A_\Phi=\Phi_0/B$ matches the unit cell area of the moir\'e bands in order to obtain the correct number of states in the bnad.  Other than
this constraint on unit cell area we can choose any two-dimensional (2D) lattice
structure.  (Here $\Phi_0=hc/e$ is the electron flux quantum.)  

When a periodic single-particle term is added
to the Hamiltonian to account for the effect of weak band dispersion,
we must choose the quasi-Bloch state lattice to match the actually lattice 
periodicity of the moir\'e system.  For
example in the case of twisted MoTe$_2$ moir{\'e}s, we must choose 
a triangular-lattice with primitive reciprocal lattice vectors $\textbf{G}_1 = (G,0)$ 
and $\textbf{G}_2 = (G/2, \sqrt3 G/2)$, where $G=4\pi/\sqrt{3}a$ 
is the lattice constant of the moir\'e reciprocal-lattice and
$a$ is the moir\'e triangular lattice constant.
In Eq.~\ref{eq:coulomb interaction}, the crystal momenta $\textbf{k}$ and $\textbf{p}$
are summed over the Brillouin zone, $\bm{q}$ is summed over unbounded momentum space, 
and the momentum labels of the quasi-Bloch states are understood to be reduced when $\textbf{p}+\textbf{q}$ or $\textbf{k}-\textbf{q}$ lie outside the
Brillouin zone (BZ) of the chosen lattice.  Explicit expressions for the form factors in Eq.~\ref{eq:coulomb interaction} and other details of the quasi-Bloch representation are provided in the appendix \ref{apx:quasi-Bloch}. 
Because of the differences between the wavefunctions of valleys with opposite Chern numbers, 
the form factors are valley-dependent and the
Hamiltonian does not possess the {\rm SU(2)} pseudospin rotational symmetry 
of the similar problem with two bands, distinguished for example by layer,
that have identical 2D band wavefunctions and the same Chern number.  
It is reduced instead to the U(1)$_v$ symmetry corresponding to valley number conservation. 

Valley polarized (VP) states are common experimentally at odd band fillings 
in moiré topological insulators and 
can be simply explained by interaction-induced spontaneous valley polarization.
Here we study the competition between these valley polarized states and 
intervalley coherent (IVC) states at odd integer filling factors. 
To be specific, intervalley coherence refers to U(1)$_v$ broken-symmetry states with non-zero values for the excitonic order parameters $\big< c^\dagger_{\uparrow\textbf{k}} c_{\downarrow\textbf{k}^\prime} \big>$.
We will focus our attention on the normal case in which the 
order parameters have $\textbf{k}=\textbf{k}^\prime$, for which
the total momentum of the condensed electron-hole-pairs is zero.

Applying Hartree-Fock mean-field theory and allowing excitonic order, we   
obtain the following expressions for the the direct (Hartree),  
\begin{equation} \label{eq:Hartree}
\begin{aligned}
    H_\text{H} =  \frac{1}{A}  \sum_{ s \textbf{k}}&
    \left(\sum_{s^\prime\textbf{p}\textbf{G}} V(\textbf{G}) \, F^2_{00}(\textbf{G})  \right.  \\
    &\left. e^{i (s \textbf{k} - s^\prime \textbf{p}) \times \textbf{G}  l_B^2} 
    \left< c^\dagger_{s^\prime  \textbf{p}} c_{s^\prime \textbf{p}} \right> \vphantom{\sum_\textbf{p}} \right) 
    c^\dagger_{s \textbf{k}} c_{s \textbf{k}},
\end{aligned}
\end{equation}
and exchange (Fock),
\begin{equation} \label{eq:Fock}
\begin{aligned}
    H_\text{F} = -& \frac{1}{A} \sum_{s s^\prime \textbf{k}} \left(\sum_{\textbf{p}\textbf{G}} V(\textbf{k}-\textbf{p}+\textbf{G}) F^2_{00}(\textbf{k}-\textbf{p}+\textbf{G}) \right. \\
    & e^{i \frac{s-s^\prime}{2} \big(\textbf{p}\times\textbf{k}+(\textbf{p}+\textbf{k})\times\textbf{G}\big) l_B^2} \left<c^\dagger_{s \textbf{p}} c_{s^\prime \textbf{p}}\right> \left. \vphantom{\sum_{\textbf{p}}} \right) c^\dagger_{s^\prime \textbf{k}} c_{s\textbf{k}}.
\end{aligned}
\end{equation}
interaction contributions to the mean-field Hamiltonian.
In the above equations, $\textbf{G}$ is summed over reciprocal lattice vectors, 
$F_{00}(\textbf{q})=\exp(-q^2l_B^2/4)$ is the form factor for $0$-th Landau level wavefunctions, 
and $l_B=\sqrt{A_\Phi/2\pi}$ is 
the magnetic length of the effective magnetic field.
The crystal momentum $\textbf{k}$ is a good quantum number in the mean-field Hamiltonian.
Given an assumed lattice structure, the
mean-field equations can be conveniently solved by diagonalizing the two-level Hamiltonian 
at each $\textbf{k}$ and densely sampling the Brillouin zone.
The formulation described here can be extended to the case in which the 
Chern bands are represented by higher Landau levels instead of the $n=0$ Landau level.
The formalism can also be extended to represent systems with several 
$|C|=1$ bands by multiple Landau levels. These extentions are 
detailed in Appendix \ref{apx: more Landau level}.

For realistic TMD systems, the bands are not perfectly flat and the form factors differ
from those of Landau levels.  To partially explore the potential consequences of 
these differences, we introduce a common 
phenomenological scaling factor $\lambda$ for the Hartree and intravalley Fock mean fields, and calculate the mean-field phase diagram 
as a function of this parameter.  Smaller $\lambda$ reduces the self-energy 
contributions that are diagonal in valley and therefore 
simulates correlation effects that prefer IVC states over VP states. 

\section{Valley Orders in Mean-Field Theory}

\subsection{Intervalley Coherent states}

Intervalley exciton condensation spontaneously breaks U(1)$_v$ symmetry.
Because the magnetic fields are opposite in the two valleys, the excitonic order 
parameter phase couples to the magnetic vector potential with charge $2e$.  
We therefore expect to find real-space vortex lattice patterns similar the Abrikosov 
those of the vortex lattices states observed in two-dimensional type-II superconductors 
in a uniform external magnetic field. The connection to the problem of superconductivity in Landau levels \cite{ tesanovic1989new,akera1991vortexlattice} can be made 
explicit by performing a particle-hole transform for one valley only, thereby mapping 
intervalley excitons to Cooper pairs and the inter-valley exchange mean-field to the BCS pairing interaction. 
The IVC ordered states, therefore, have two vortices with the same vorticity in each unit cell, breaking the continuous translation symmetry to a lattice symmetry.
Our mean-field calculations confirm that the lowest energy mean-field state in the absence of an external potential, {\it i.e.} in the pure Landau level pairing problem, is a triangular vortex lattice state. 
However, we find that this state is incompatible with the triangular single-particle moir\'e lattice and instead appears when we start from a rectangular magnetic lattice, as detailed in an Appendix \ref{apx:SC VL}.  
Since the band dispersion and non-ideal geometry in the real moir\'e materials 
can lower the energy of states with compatible periods and the different 
vortex lattice states are quite similar in energy, we do not discuss these solutions at length.


\begin{table}
    \centering
    \begin{tabular}{|c|c|}
        \hline
        Symmetries & Energy per unit cell ($e^2/\epsilon\, l_B$)\\
        \hline
        C$_{3z}$ & -0.175825  \\
        C$_{3z}+4\pi/3$ & -0.167919  \\
        Trans       & -0.174503  \\
        \hline
        ATVL & -0.180578 \\
        \hline
        VP & -0.565462 \\
        \hline
    \end{tabular}
    \caption{Energies of self-consistent mean-field solutions with different crystal symmetries, as explained in the main text. 
    The moir\'e lattice calculations were performed using
    a 60-by-60 momentum-space grid, with gate distance $D$=20nm and magnetic length $l_B$$\approx$3.57nm, and under the lowest-Landau-level approximation.
    Energies of all opposite Chern number IVC state solutions are computed with $\lambda$=0 so that only the intervalley exchange is included and both inversion and time-reversal symmetries are conserved. 
    The C$_{3z}$, C$_{3z}+4\pi/3$, and Trans states, exhibiting different crystal symmetries as explained in Appendix \ref{apx:crystal symmetry}, are self-consistent solutions with the moir\'e period.
    The Abrikosov triangular vortex lattice solution (ATVL) is obtained when the 
    single-particle lattice is replaced by rectangular lattice with the same unit-cell area, as detailed in the Appendix \ref{apx:SC VL}.
    Because its lattice is incompatible with the triangular moir\'e lattice, it is unlikely to manifest lowest energy in the real system.
    The energy of the valley polarized state, calculated with full strength 
    intravalley exchange interactions, is also listed for comparison.
    Because the intervalley coherent states have half-filling in each valley, 
    their energies would be half of the valley polarized energy if 
    intervalley and intravalley exchange where of equal strength.}
    \label{tab:crystal symmetry}
\end{table}

The Hartree-Fock ground state is a single-Slater-determinant 
state with a single state occupied at each $\bm{k}$ in the BZ that 
is a coherent superposition of opposite valleys.  The $\bm{k}$-dependent 
occupied states can be parametrized by the valley Bloch sphere polar and azimuthal 
angles $\theta_\textbf{k}$ and $\phi_\textbf{k}$; $\phi_\textbf{k}$ is the phase of 
the non-zero order parameter 
$\big< c^\dagger_{\uparrow\textbf{k}} c_{\downarrow\textbf{k}} \big>$,
and $\theta_\textbf{k}\in[0,\pi]$ is determined by the valley/spin polarization $S_z = \cos\theta_\textbf{k} = 2\big< c^\dagger_{\uparrow\textbf{k}} c_{\uparrow\textbf{k}} \big>-1$.  Whereas $\theta_\textbf{k}$ is periodic across the BZ,
\begin{equation} \label{eq:phi plus G}
    \phi_{\textbf{k}+\textbf{G}} = \phi_\textbf{k} +  (\textbf{G}\times\textbf{k})_z\ l_B^2, 
\end{equation}
due to the due to the Chern number mismatch.
A direct consequence is that the winding number of $\phi_\textbf{k}$ 
around the Brillouin zone is 2.
A general conclusion is derived in the Appendix \ref{apx:phase winding number} 
that the winding number of the coherence phase $\phi_\textbf{k}$
between two bands is equal to the Chern number difference. 
For systems with no Chern number mismatch between bands, 
IVC states usually have constant $\phi_\textbf{k}$ and no valley polarization ($\theta_\textbf{k}=\pi/2$), and a delicate competition between VP and IVC states
\cite{eisenstein2004bose,jungwirth2000pseudospin}. 
The situation is different for moir{\'e} topological insulators, 
because of the Dirac cones induced by the winding of $\phi_\textbf{k}$, and because of 
the role of inversion and time-reversal symmetry breaking on which we elaborate below.

\begin{figure*}
    \centering
    \includegraphics[width=\linewidth]{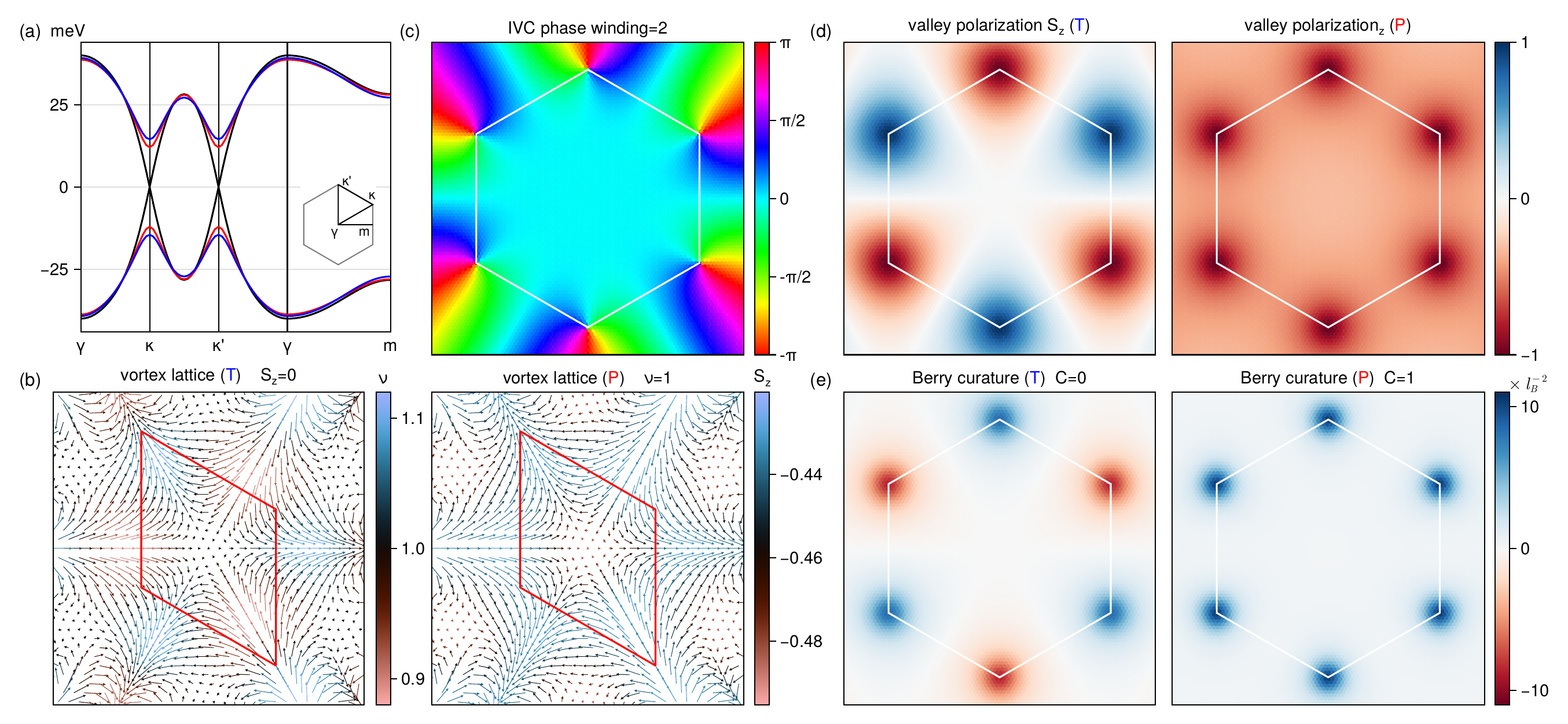}
    \caption{
    Hartree-Fock self-consistent intervalley coherent (IVC) states with C$_{3z}$ rotational symmetry have two Dirac points located at the two inequivalent 
    corners of the moir{\'e} Brillouin zone.
    (a) Gapless bands of IVC states with both 2D inversion symmetry (P) and time reversal symmetry (T) plotted as solid black lines.
    Bands are also plotted as red (blue) lines for self-consistent solutions 
    with P (T) symmetry preserved but T (P) broken. 
    (b) Arrows show the real-space distribution of the intervalley coherence pseudospin
    $S_{x}-iS_y= \langle c^\dagger_{\uparrow\textbf{k}} c_{\downarrow\textbf{k}^\prime} \rangle $, revealing its vortex lattice.  In the moir{\'e} unit cell outlined
    by the red parallelogram, two vortices merge into a texture with vorticity $2$.
    For T symmetric states, the valley polarization $S_z$=0 everywhere 
    but the local charge density $\nu$ varies as illustrated by the color scale;
    For P symmetric states, the density is uniform ($\nu=1$ everwhere) but 
    the spin-polarization $S_z$ varies as illustrated by the $S_z$ color scale. 
    (c,d,e) momentum-space plots in which the Brillouin zone is outlined by white hexagons.
    (c) the coherence phase $\phi_k$ winds twice in each reciprocal unit cell, as explained in the main text.
    The winding centers are the band-crossing Dirac points $\kappa$ and $\kappa^\prime$.
    This distribution is nearly the same for the three different IVC states.
    (d) The P/T gapped IVC states have different valley polarization behavior
    near the two Dirac points. 
    Electrons are polarized to opposite valleys at $\kappa$ and $\kappa'$
    in the T gapped states and to the same valley in the P gapped states.
    (e) The Berry curvature distributions are related to those of the valley polarization.
    The T gapped states have Chern number 0 and the P gapped states have Chern numbers $\pm$1 depending on the sign of valley polarization.
    }
    \label{fig:C3}
\end{figure*}

Eq.~\ref{eq:phi plus G} requires that
at least two phase-winding singularities be present in the Brillouin zone.
Other symmetries impose further requirements that fix the number and positions of phase singularities. 
Derivation of the corresponding symmetry transforms and results are detailed in the Appendix \ref{apx:symmetry} and also summarized here.
Any solutions to the Hartree-Fock theory introduced in last section must have the lattice translation symmetry defined together with the corresponding quasi-Bloch representation.
Besides, it can have the in-plane inversion (P/C$_{2z}$) and the time-reversal (T) symmetry.
Both symmetry requires $\phi_{-\textbf{k}}=\phi_\textbf{k}$, but have distinct requirements on $\theta$.
P symmetry requires $\theta_{-\textbf{k}}=\theta_\textbf{k}$, while T symmetry requires $\theta_{-\textbf{k}} = \pi-\theta_\textbf{k}$. 
Moreover, the solution can exhibit different crystal symmetries.
It can be invariant under threefold rotation (C$_{3z}$) or this rotation composed of a phase addition (C$_{3z}+4\pi/3$); or it can exhibit half-lattice translation symmetry (Trans) along $G_1/2$ (or its rotational equivalents).
As explained in the Appendix, these are the only possible symmetries that allow only two phase singularities.
We calculate the self-consistent IVC solutions with these symmetries at $\lambda=0$ and list their energies in Table ~\ref{tab:crystal symmetry}.
The energies of the triangular Abrikosov's vortex lattice and VP states with $\lambda=1$ are also listed as references.
Note that the energies of IVC states are similar for different vortex lattice arrangements, including the triangular vortex lattice state, and that the intervalley interaction is weaker than the intravalley interaction, as all IVC energies in magnitude are smaller than half of the VP energy. (Half because in IVC states each valley has only half filling.)

The C$_{3z}$ solution with the two singularities located
at the moir{\'e} Brillouin zone corners $\kappa$ and $\kappa^\prime$, as shown in Fig.~\ref{fig:C3}, has the lowest energy among solutions in compatible with moir\'e lattice symmetry, and our discussion in the next section will be focused on this state.

\begin{figure*}
    \centering
    \includegraphics[width=\linewidth]{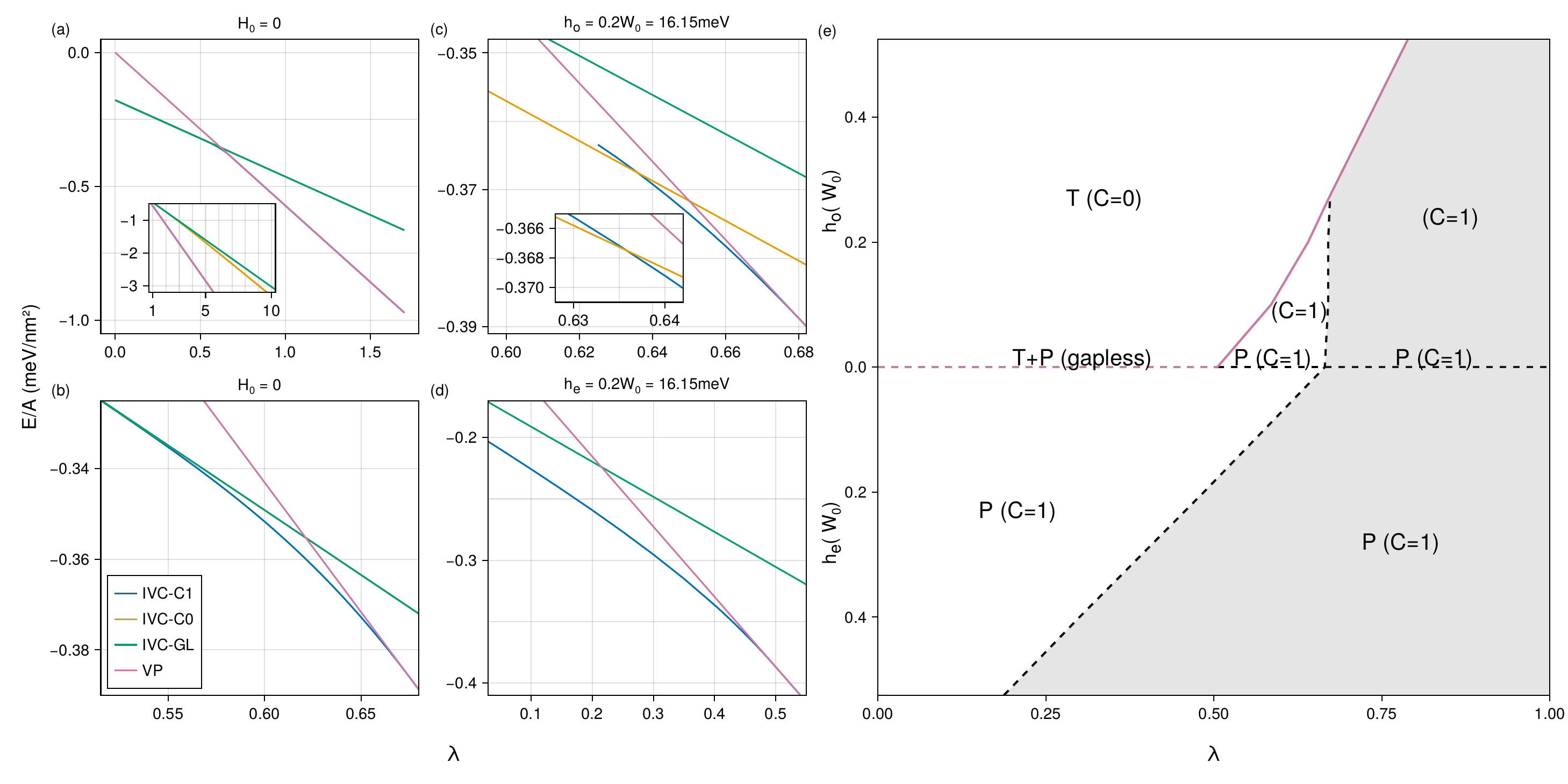}
    \caption{
    Phase diagrams {\it vs.} the intralayer interaction strength parameter $\lambda$ 
    and the band dispersion term $H_0$ in equation \ref{eq:H0 T} or \ref{eq:H0 P} include four distinct states.  In addition to valley polarization (VP) phase and gapless
    intervalley coherence (IVC-GL), states are disinguished by the absence or presence of valley polarization at the two Dirac points.  Gapped states can be topologically
    trival (IVC-C0) or have non-zero Chern numbers $\pm1$ (IVC-C1).
    (a) and (b) compare the energies of these four typical phases with $H_0=0$. 
    With increasing $\lambda$, the ground state first experiences a second order phase transition in which gaps open time-reversal symmetry is broken, followed by a second phase transition into a VP state as the coherence gradually decreases to zero. 
    This two-step phase transition circumvents the first order phase transition directly from a gapless IVC state to a VP state that is typical when the bands are not 
    topological. A state with a spontaneous gap that breaks 2D inversion symmetry is metastable only at higher $\lambda$ and never becomes the ground state.
    (c) shows energies when a time-reversal-invariant band dispersion, 
    relevant in realistic moir{\'e} materials, is added to the Hamiltonian. 
    Such a term lowers the energy of IVC-C0 phase, making it the ground state in place of 
    the IVC-GL phase over a small $\lambda$ range.
    The transition from the IVC-C0 to the IVC-C1 phase 
    is first-order when $H_0$ is small, as shown in the plot, and is 
    followed by a transiton to a VP phase as in the previous case.
    When $H_0$ becomes large, the transition is directly to the VP phase.
    (d) shows the energies for the case in which a time-reversal-symmetry-breaking $H_0$ is present.  This type of term does not appear in real physical systems.
    The ground state in this case is either in the IVC-C1 phase for small $\lambda$ or in the VP phase for large $\lambda$.
    (e) Phase diagram {\it vs.} $\lambda$ and $H_0$: 
    The upper half is for the time-reversal-invariant $H_0$ and the lower half is for the time-reversal-symmetry-breaking $H_0$.
    The grey regions are the VP phase and the white regions are different IVC phases;
    preserved symmetries (T=time-reversal and P=inversion)
    and Chern numbers are indicated for all regions.
    Solid boundary lines stand for first order phase transition and 
    dashed lines stand for second order.
    The purple lines indicate topological phase transitions accompanied by a change in Chern number.  The parameters $h_{o}$ and $h_{e}$ are defined in the main text.
    }
    \label{fig: energy lines and phase diagram}
\end{figure*}

\subsection{Valley Order Competition}

At each momentum-space phase singularity point, 
the mean-field eigenvalues can touch or split with the 
occupied level polarized to one valley or the other.  These two 
cases correspond, respectively, to gapless and gapped Dirac cones.
The two Dirac cones have the same phase chirality.
If both T and P are conserved, $\theta=\pi/2$ at both Dirac points ($\textbf{k}_D$)
and both are gapless. Gaps can be opened by spontaneously breaking the 
P or T symmetries or by adding 
an external potential $H_0 = \sum_\textbf{k}h_\textbf{k} \sigma_{z\textbf{k}}$,
where $\sigma_{z\textbf{k}}$ is the valley Pauli matrix;
valley $\sigma_{x,y}$ terms are forbidden by the U(1)$_v$ valley symmetry.
The origin of $H_0$ can be simple band dispersion or external Zeeman fields.
If $h_\textbf{k}$ is odd in $\textbf{k}$, $H_0$ conserves T but breaks P.  
As an example of a P breaking single-particle term we consider the model,
\begin{equation} \label{eq:H0 T}
    h_\textbf{k} = h_o \big[\sin(\textbf{k}\cdot\textbf{a}_1) + \sin(\textbf{k}\cdot\textbf{a}_2) + \sin(\textbf{k}\cdot\textbf{a}_3) \big],
\end{equation}
where $a_{1,2,3}$ are the three smallest non-zero lattice vectors;
On the other hand contrary, if $h_\textbf{k}$ is even, $H_0$ conserves P but breaks T.
As an example of a T breaking single-particle Hamiltonian we consider,
\begin{equation} \label{eq:H0 P}
    h_\textbf{k} = h_e.
\end{equation}
The value of $h_{o/e}$ controls the gap size.
In Fig.~\ref{fig:C3}(a), the bands of C$_{3z}$ IVC states with P/T/P+T symmetries are plotted with red/blue/black lines.
The gapless bands are manifested by two interaction-induced Dirac cones.

The arrow plots of Fig.~\ref{fig:C3}(b) show the vortex 
lattice pattern of local valley-pseudospin polarizations in real space. 
Due to the effective magnetic fields, the distributions are not periodic in the unit 
cells plotted as the red parallelograms.
If the T symmetry is preserved, the z-component of valley pseudospin is zero everywhere, but the electron number density can vary due to broken P symmetry.
On the other hand if P symmetry is preserved but T symmetry is broken, $S_z$ fluctuates and the local $\nu$ is constant.
Fig.~\ref{fig:C3}(c) shows the winding of the 
coherence phase $\phi_\textbf{k}$ as a continuous function of momentum, where the white hexagon denotes the first Brillouin zone, and the two winding centers are the two Dirac points at Brillouin zone corners.
Due to U(1)$_v$ symmetry, global increment of $\phi_\textbf{k}$ connects different symmetry-broken ground states.
The $\phi_\textbf{k}$ shown here is obtained with $\phi_{\textbf{k}=0}=0$.
This pattern is universal for all IVC states with a given symmetry.

When a gap is opened at the Dirac points, the occupied band is 
polarized to one valley and can have Berry curvatures and Chern numbers. 
T symmetry requires that the states at the two Dirac points 
be polarized to opposite valleys, 
while P symmetry requires them to be polarized to the same valley.
Due to the property that the two Dirac cones share the same chirality, 
bands in the former case have Chern number 0, and in the latter case have Chern number $\pm1$.  The details of the Berry curvature calculations whose results are 
summarized in Fig.~\ref{fig:C3} are discussed in the Appendix \ref{apx: Berry}.

Fig.~\ref{fig: energy lines and phase diagram} plots the energies of competing states as a function of model parameters.  The subfigures (a) and (b) show how P and T symmetries are 
broken spontaneously without $H_0$.  The gapless IVC phase is the ground state at $\lambda=0$.  As $\lambda$ increases, the energies of VP phases keep decreasing and 
eventually become lowest in energy. 
However, near the crossing point between the energies of
gapless IVC and VP states, a gapped IVC phase with a nontrivial Chern number and broken T symmetry emerges as the ground state.
The system transforms between IVC and VP states via 
two second-order phase transitions; a nonzero gap develops first and the intervalley 
coherence then gradually decreases to zero.
A spontaneous gapped phase with T symmetry and zero Chern number is never the ground state, despite having lower energy than the gapless IVC state at very high $\lambda$.
However, when a band dispersion therm that explicitly breaks P symmetry is 
introduced via Eq.~\ref{eq:H0 T}, the energy of this state is lowered, and it becomes the ground state for small $\lambda$, replacing the gapless IVC phase.
The gapped IVC state with non-trivial Chern number survives 
as a metastable state at small $H_0$.
The two Dirac points are polarized to the same valley but the two
gaps are different, and neither P nor T symmetry is preserved.
This state has a first-order topological phase boundary with the trivial Chern band state, as show in subfigure (c); when $H_0$ is large enough, it is never the ground state.
On the other hand, if $H_0$ is a Zeeman-like term as in Eq.~\ref{eq:H0 P},
the non-trivial IVC state is the ground state of a large range of small $\lambda$, as shown in Fig.~\ref{fig: energy lines and phase diagram}(d).

All of these phases and the transitions between them
are summarized in Fig.~\ref{fig: energy lines and phase diagram}(e). 
The horizontal axis is $\lambda$ and the vertical axis is the band dispersion, with
the non-dispersive case lying in the center, and the top and bottom halves for dispersive bands respecting T and P symmetries, respectively.
The symmetries and Chern number of each phase are specified in the plot.
The white regions are phases with nonzero intervalley coherence, while the grey regions are VP phases with Chern number $\pm 1$.
Phase transitions between phases with different Chern numbers are topological and are plotted with purple boundary lines.  Non-topological boundaries are plotted as black lines,
solid phase boundary lines denote first-order phase transitions, and the dashed lines correspond to second-order phase transitions.
The IVC phase region expands with band dispersion with time-reversal symmetry and can even reach $\lambda = 1$ with large enough dispersion exceeding the plotting range, which is unphysical for tMoTe$_2$ systems.

\section{Magnetic-Field-Dependent Many-Body Insulators}

Magnetic fields enter the Hamiltonians of non-relativistic two-dimensional 
systems in two distinct ways, first orbitally by replacing momentum $\bm{p}$ by $\bm{p}+(e/c)\bm{A}$, where $\bm{A}$ is the magnetic vector potential, and second through Zeeman coupling to spin.  The discussion in this section, which contrasts the 
fractional quantum anomalous Hall effect (FQAHE) and the fractional quantum spin 
Hall effect (FQSHE), requires that these two couplings be distinguished and treated as separately controllable.  

Twisted MoTe$_2$ moir\'es and rhombohedral graphene multilayers are at present the only physical systems\cite{cai2023signatures, zeng2023thermodynamic, park2023observation, xu2023observation, lu2024fractional} in which the fractional quantum anomalous Hall effect (FQAHE) - a fractional Hall effect in the absence of a magnetic field - has been observed.
Microscopically, the QAHE is a property \cite{macdonald1994introduction, mac1995akkermans} of 2D insulators
with a gap that appears at a density $n^*$ that depends on the orbital magnetic field.
The quantized Hall conductivity $\sigma_{H}$ 
is related to $n^* = N^*/A$ by the Streda formula: \cite{streda1982theory,widom1982thermodynamic},
\begin{equation}
\frac{\sigma_H}{e^2/h} = \frac{dN^*}{dN_{\phi}},
\end{equation}
where $A$ is the system area and $N_{\phi}=AB/\Phi_0$ is the number of magnetic flux 
quantum penetrating the 2D system.  For example, the ordinary integer and fractional 
quantum Hall effects appear respectively at integer and fractional Landau level filling 
factors $\nu_{LL}=N/N_{\phi}$, and therefore at densities
$n^*=\nu_{LL} N_{\phi}/A$ that are linear in field $B$. As we explain below, the QSHE
occurs when there are separate spectral gaps for spin-$\uparrow$ and spin-$\downarrow$,
electrons with opposite dependences of critical density on magnetic field.
( Similar arguments have been advanced previously \cite{yang2006stvreda,monaco2021stvreda}.) 

\begin{figure}[t]
\includegraphics[width=\linewidth]{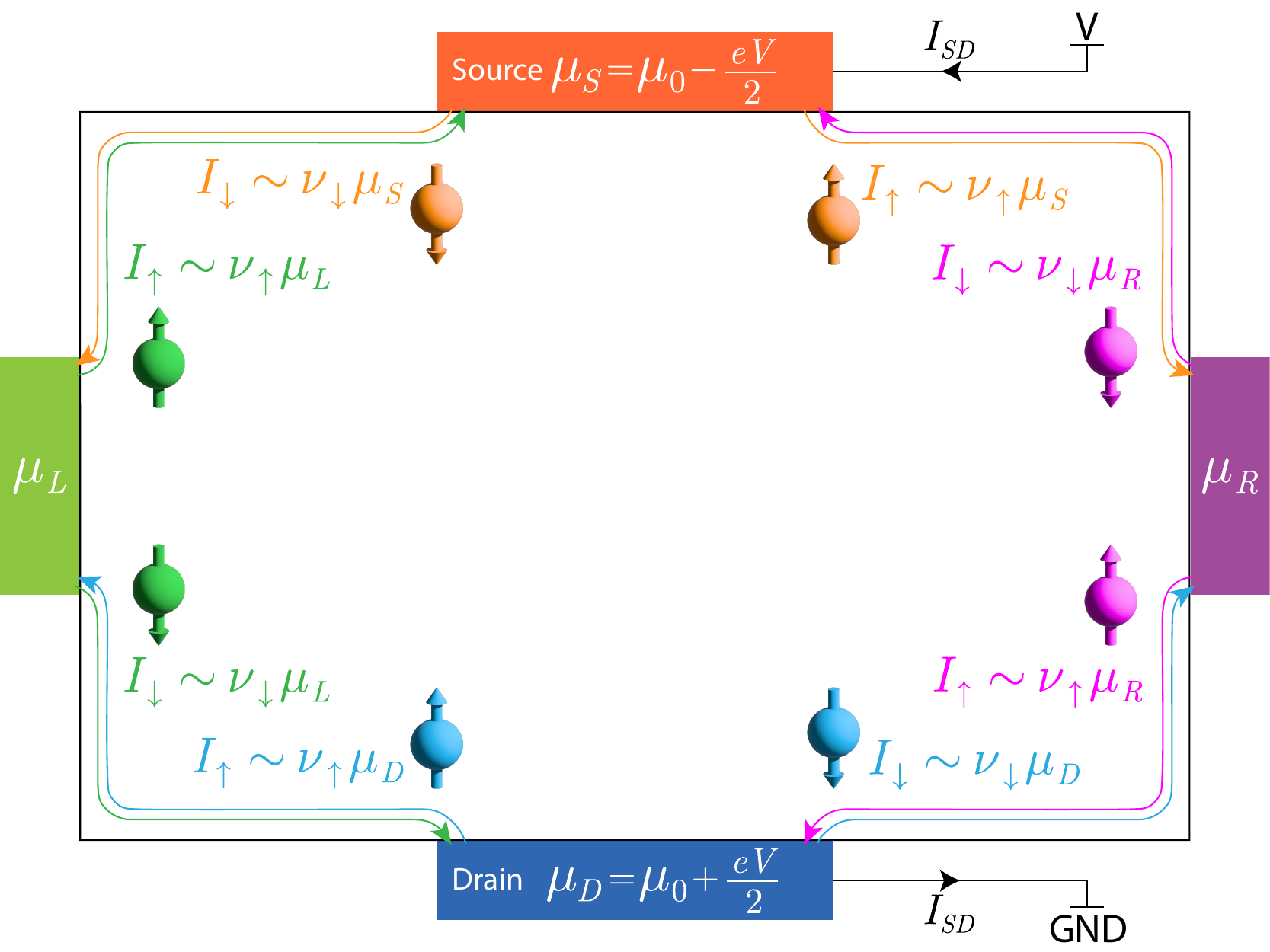}
\caption{Quantum Spin Hall Effect.  Consider a 2D system at band filling 
$\nu=2$ with conserved charge and spin that has a band gap for both spins.  When the chemical potential for spin $s$
changes within the gap it induces an edge current which satisfies $dI_{e}/d\mu_s=(e/h) C_s$
where $C_s$ is the Chern number for band $s$.  The current $I_e$
flowing along each edge segment is $ (e/h) C_{\uparrow} (\mu_{\uparrow}-\mu_{\downarrow})$
where $\mu_{s}$ is the chemical potential of the contact of that edge segment for 
spin $s$.  Gapped IVC states can establish an arbitrary 
chemical potential difference between spins in the bulk by letting the global
phase difference between spins be time dependent: $\hbar \dot{\phi} = \mu_{\uparrow} 
\mu_{\downarrow}$.  The absence of a bulk spin gap seems to forbid a QSHE.
The illustration shows a typical setup in QSHE experiments with two extra contacts for the purpose of Hall conductance mearsurement. 
}
\label{Fig:Schematic}
\end{figure}

The following argument parallels one that can be used to derive the Streda formula.  
We assume a finite area 2D electron system in which the number of 
electrons $N_s$ in the two spins/valleys are separately good quantum numbers.
It is therefore possible to define separate chemical potentials for the two-spins.
If the 2D electron system is an insulator, it must be an insulator for both spins - that 
is to say that there is an interval of energy which is inside the gap for both spins.
This is what happens at all even integer band filling factors, for example, and it gives rise to a spin Hall effect manifested by the regular non-local transport signals.
When the chemical potentials lie in these gaps, a response of any physical observable to a change in chemical potential within the gap can be produced only at the edge.
In particular, any change in orbital magnetization $M_{orb}$ must come from currents $I_e$ that flow at the edge.  It follows that 
\begin{equation}
\label{eq:spinstreda}
\frac{\partial I_e}{\partial \mu_s}= \frac{c}{A} \frac{\partial M_{orb}}{\partial \mu_s} = \frac{e}{h} \frac{\partial N_s}{\partial N_{\phi}},
\end{equation}
where $N_{\phi}=AB/\Phi_0$.  
The first equality in Eq.~\ref{eq:spinstreda} can be obtained by defining the orbital 
magnetization operator in terms of the derivative of the Hamiltonian with respect to the 
magnetic field that appears in $\bm{A}$ and the second is
a thermodynamic Maxwell relation.
At band filling factor $\nu=2$, both bands are full and  
\begin{equation}
\frac{\partial N_s}{\partial N_{\phi}} = \pm C_s,  
\end{equation}
where $C_s$ is the Chern number of band $s$, since we know that the Streda formula applies
separately to the two bands.  Since spin/valley is conserved in the system, differences 
in chemical potential can only be relaxed in the contacts.  Along a segment of the edge between two contacts the upstream and downstream contacts will act as reservoirs for $\uparrow$ and $\downarrow$ particles, which can therefore have different chemical potentials $\mu_{\uparrow}$ and $\mu_{\downarrow}$.
The current flowing along a given segement of the edge is 
\begin{equation}
I_{e}= \frac{e}{h} C_{\uparrow} (\mu_{\uparrow}-\mu_{\downarrow}),
\end{equation}
where $\mu_{s}$ is the chemical potential of the contact that is upstream of that 
edge segment for spin $s$.  (See Fig.~\ref{Fig:Schematic}.)   

Gapped valley-polarized states are inconsistent with a fractional quantum spin Hall effect 
because they are necessarily accompanied by a quantum anomalous Hall effect.
Can gapped IVC states explain the fractional quantum spin Hall effect phenomenology  
uncovered in reference \cite{kang2024evidence}.  The observations in \cite{kang2024evidence}
can be fully explained by a bulk state with dissipationless edge currents  
$\partial I_e/\partial \mu_s = s e/2h$ for spin $s$.
By the argument outlined above, these edge state properties are 
supported by bulk insulators with $\partial (N_{\uparrow}-N_{\downarrow})/\partial N_{\phi}=1$ - half as large as in the case of a fully filled band.  
Intervalley coherence on its own cannot explain this behavior since IVC states are chargeinsulators but have gapless spin-excitations. There are solutions to the bulk
time-dependent mean-field equations with a time-dependent global phase difference 
corresponding to a chemical potential difference between spins
$\hbar \dot{\phi} = \mu_{\uparrow} - \mu_{\downarrow}$ for any value of $\hbar \dot{\phi}$.
The time-dependent phase gives rise to an effective Zeeman coupling $h_{e}$
that produces a finite spin polarization that does not have a specific
relation to the orbital magnetic field.  Although bulk IVC states do not explain 
the FQSHE, they could still be consistent with its observation if intervalley coherence
was somehow absent near the edge of the sample.

\section{Discussion}

In this article we have examined the stability of inter-valley coherent insulating 
states of moir\'e topological insulators, {\it i.e.} moir\'e materials in which 
opposite valleys have non-zero Chern numbers of opposite sign, at odd integer 
filling factors.  $t$MoTe$_2$ and $t$WSe$_2$ moir\'e materials are examples   \cite{wu2019topological} of moir\'e topological insulators.  
In these materials, spin is locked to valley by strong spin-orbit interactions 
and the number of $\uparrow$ and $\downarrow$ electrons ($N_{\uparrow}$ and $N_{\downarrow}$ are conserved separately, and inter-valley coherence breaks
this symmetry.  Our explicit calculations were performed using 
Landau levels with opposite signs of magnetic field 
as a generic representation of the Chern bands in the limit in which 
their band widths are negligible and correlations are strong.
At odd integer filling factors the active Chern band is  
half-filled and insulating ground states usually arise from broken symmetries.
One important finding from our work is that valley polarized insulating states 
are always lower in energy than inter-valley coherent states when particles interact
by Coulomb interactions.  In order to obtain inter-valley coherent states
we scale intra-valley interactions down by a factor of $\lambda < 1$.
Possible origins of such a scaling are discussed below. 
This scaling favors inter-valley coherent interactions since they are stabilized
by inter-valley exchange.  The valley-coherent states have two interaction-induced 
momentum-space Dirac cones with the same chirality, that can be gapped either 
spontaneously by breaking time-reversal or inversion symmetry
or by adding external fields that account for band dispersion effects.
These conclusions also apply to Chern bands represented by higher Landau levels.

Our work was motivated in part by the observation \cite{kang2024observation} of a fractional quantum spin Hall effect in $t$MoTe$_2$ at a small relative twist angle.
It is natural to consider whether or not inter-valley coherence in some devices could explain this behavior.  Many-body states that have gaps for both charge and spin 
can be characterized by the derivatives 
\begin{equation}
\label{eq:stredaspin}
    \sigma_{s} = \frac{\partial N_{s}}{\partial N_\phi}.
\end{equation}
In Eq.~\ref{eq:stredaspin} $N_{\phi}$ is the number of 
flux quanta passing through the 2D system and $p$ is the separation in flux quanta states 
that are free of quasiparticle excitations.  Since the FQSHE state is time-reversal 
invariant $\sigma_{\uparrow}=-\sigma_{\downarrow}$.  The transport properties associated 
with the FQSHE appear when $\sigma_{\uparrow}=-\sigma_{\downarrow}=1/2$.
Inter-valley coherence cannot on its own explain the FQSHE because these states are 
do not have a gap for spin-excitations.  However, it could still play a role in 
explaining this behavior by providing a bulk charge gap.  In that case 
intervalley coherence would have to be absent at the edge of the system, 
perhaps by forming separate domains of spin-polarized states that are 
aligned along the edge.  The final explanation for the FQSHE observations 
awaits further work.

If intervalley-coherent states do not occur in single-bilayer moir\'e topological 
insulators, how can they be stabilized?  One possibility is to arrange for the 
two-valleys to occupy different layers in a 2D crystal stack.  This change makes 
inter-valley interactions weaker than intra-valley interactions, in contrast to the 
$\lambda < 1$ case considered in our explicit calculations.  However, layer separation
introduces a capacitive Hartree-energy cost \cite{jungwirth2000pseudospin} for valley polarization and creates a non-trivial competition between valley-coherent states and 
stripe-domain valley-polarized states.  Since domain walls between regions with opposite
senses of valley polarization, and therefore opposite Chern number sign, carry 
chiral currents, layer separation between Chern bands should prove to be an interesting 
tuning variable to explore new types of strongly interacting topological states.

\section{Acknowledgements}
We thank Nemin Wei, Tobias Wolf, Michael Zaletel, Taige Wang, Daniel Parker, Kin Fai Mak, Jie Shan, and Yves Kwan for helpful discussions.  This work was 
supported Simons Foundation Targeted Grant Award No. 896630.

\newpage
\appendix
\onecolumngrid
\section{Introduction to Magnetic Quasi-Bloch representation} \label{apx:quasi-Bloch}

The energy spectrum of a two-dimensional electron gas in a uniform perpendicular magnetic field consists of highly degenerate Landau levels. 
States in the same Landau level are usually distinguished by guiding centers or angular momenta, but can be labeled instead by quasi-Bloch representation.  We choose 
that representation in our paper to highlight the role of Landau levels as a convenient 
generic stand-in for flat Chern bands.  The eigenstates of the magnetic translation operator on a quasi-periodic lattice with one flux quantum per unit cell
form a complete basis of the Hilbert subspace of every Landau level, and each eigenstate is associated with a crystal momentum in the corresponding Brillouin zone.
When periodic boundary condition are applied to a finite-size region, 
the crystal momentum is sampled on a discrete mesh.
This idea has been developed previously in Refs. \cite{brown1964bloch, zak1964magnetic, haldane1985many, haldane2018origin}.
This appendix provides a brief introduction to quasi-Bloch states and presents some important results that are useful in the main text.
$\hbar=c=1$ is assumed for convenience in this section.

\subsection{Magnetic Translation Operator}

The uniform magnetic field is in the $\pm z$ directions.
We use the symbol $\sigma = \pm 1$ to denote its direction and $B$ for its magnitude.
\begin{equation}
    \vec{B} = -\sigma B \hat{z}.
\end{equation}
The magnetic length $l_B = 1/\sqrt{eB}$. 
Classical electrons with charge $-e$ rotate in the $x$-$y$ plane with cyclotron frequency
$\boldsymbol{\omega_c} = -\sigma (eB/m) \hat{z}$ and position can be separated into 
orbit-center $C$ and cyclotron-orbit $\boldsymbol{\rho}$ contributions: 
\begin{equation}
    \mathbf{r} = \mathbf{C} + \boldsymbol{\rho},
\end{equation}
where $\mathbf{C} = (X, Y)$ is the orbit-center, and $\boldsymbol{\rho}$ is the rotational movement. The kinetic momentum $\boldsymbol{\Pi} = m\textbf{v} = \textbf{p}+e\textbf{A}$, where the velocity comes from the rotation,
$\textbf{v}=\boldsymbol{\omega_c}\times\boldsymbol{\rho}$.  It follows that 
\begin{equation}
    \boldsymbol{\rho} = -\frac{\boldsymbol{\omega_c}\times\textbf{v}}{\omega_c^2} = \sigma \hat{z}\times(\textbf{p}+e\textbf{A})l_B^2,
\end{equation}
and hence that 
\begin{equation} \label{eq:guidingcenter}
    \left\{
    \begin{aligned}
        X &= x + \sigma\,\Pi_y\,l_B^2\,,\\
        Y &= y - \sigma\,\Pi_x\,l_B^2\,.
    \end{aligned}
    \right.
\end{equation}
These guiding center coordinates are conserved and equivalent to the 
the conserved magnetic momentum
$\tilde{\textbf{p}}=\sigma\hat{z}\times\textbf{C}\,l_B^{-2} =  \textbf{p}+e\textbf{A} + \sigma\hat{z}\times\textbf{r}\,l_B^{-2}$:
\begin{equation}
    \left\{\,
    \begin{aligned}
        \tilde{p}_x &= -\sigma\,Y/l_B^2 = p_x + eA_x -\sigma\, y/l_B^{2}\,,\\
        \tilde{p}_y &= +\sigma\,X/l_B^2 = p_y + eA_y +\sigma\, x/l_B^{2}\,. 
    \end{aligned}
    \right.
\end{equation}

Quantum mechanics can be invoked by recognizing that
position $\textbf{r}$ and $\textbf{p}$ are canonical operators.
The magnetic translation operator
\begin{equation}
    t(\textbf{a}) = e^{-i\tilde{\textbf{p}}\cdot\textbf{a}},
\end{equation}
and commutes with the Hamiltonian.
When the magnetic field is decreased to zero from either direction, $\tilde{\textbf{p}}$ reduces to $\textbf{p}$ and the magnetic translation operator reduces
to the normal translation operator $T(\textbf{a}) = e^{-i{\textbf{p}}\cdot\textbf{a}}$.

We compute a series of useful commutators between these operators.
For the kinetic momentum,
\begin{equation}
    [\Pi_x, x] = [\Pi_y, y] = -i,\ [\Pi_x, y] = [\Pi_y, x] = 0,\ [\Pi_x, \Pi_y] = i\sigma eB = {i\sigma}/{l_B^2}.
\end{equation}
Defining
\begin{equation}
    a^\dagger = \frac{(\Pi_x - i\sigma\Pi_y)l_B}{\sqrt{2}},\ a = \frac{(\Pi_x + i\sigma\Pi_y)l_B}{\sqrt{2}}
\end{equation}
so that $[a,a^\dagger]=1$, the Hamiltonian 
\begin{equation}
    H = \frac{1}{2m}(\Pi_x^2 + \Pi_y^2) = |\omega_c| \big(a^\dagger a + \frac{1}{2}\big).
\end{equation}
Next we calculate a set of useful commutators involving guiding 
center $\textbf{C}$ or $\tilde{\textbf{p}}$:
\begin{equation} \label{eq:commutators}
\left\{ \ 
\begin{aligned}
    &[\Pi_x, \tilde{p_x}] = [\Pi_y, \tilde{p_x}] = [\Pi_x, \tilde{p_y}] = [\Pi_y, \tilde{p_y}] = 0,\\
    &[\tilde{p_x},\tilde{p_y}]=-i\sigma / l_B^2,\\
    &[H, \tilde{p_x}] =[H,\tilde{p_y}]=0,\\
    &[H,t(\textbf{a})]=0.\\
\end{aligned} 
\right.
\end{equation}
It follows that the operators 
\begin{equation}
    b^\dagger = \frac{(\tilde{p}_x - i\sigma \tilde{p}_y)l_B}{\sqrt{2}},\ b = \frac{(\tilde{p}_x + i\sigma \tilde{p}_y)l_B}{\sqrt{2}}
\end{equation}
with commutator $[b,b^\dagger]=1$.
commute with $a^\dagger$, $a$ and $H$.

The Landau levels are defined by the eigenvalues of $a^\dagger a$;
in order to distinguish the states within the same level, we need to use the ``$b$" - type operators.  Two common choices are the guiding center $X$ representation, and the angular momentum representation of eigenstates of $b^\dagger b$.
The magnetic translation operators provide an alternate choice, 
the quasi-Bloch representation, on which we now focus.

\subsection{Magnetic Quasi-Bloch States} 

Textbook Bloch wavefunctions are eigenstates of lattice translation operators
in absence of the magnetic field.  The presence of a magnetic field modifies the 
translation operators and the way in which periodic boundary conditions are applied 
to finite-size systems.
In a magnetic field distinct magnetic translation operators usually do not commute. 
Using $[\tilde{\textbf{p}}\cdot\textbf{a}_1,  \tilde{\textbf{p}}\cdot\textbf{a}_2] = -i\sigma(\textbf{a}_1\times\textbf{a}_2)\cdot\hat{z}/l_B^2$ and the Baker–Campbell–Hausdorff formula yields 
\begin{equation}\label{eq:translation commutator}
    t(\textbf{a}_1) t(\textbf{a}_2) = t(\textbf{a}_2) t(\textbf{a}_1) e^{i\sigma(\textbf{a}_1\times\textbf{a}_2)_z/l_B^2} = t(\textbf{a}_1+\textbf{a}_2) e^{i\frac{1}{2}\sigma(\textbf{a}_1\times\textbf{a}_2)_z/l_B^2}.
\end{equation}
(Here the cross product of two in-plane vectors with a subscript $z$ denotes
for its $z$ component: $(\textbf{a}_1\times\textbf{a}_2)_z = (\textbf{a}_1\times\textbf{a}_2)\cdot\hat{z}$.)
However, consider a lattice generated by primitive vectors
$\textbf{a}_1$ and $\textbf{a}_2$ that satisfy
\begin{equation}
    \textbf{a}_1\times\textbf{a}_2 = 2\pi l_B^2 \hat{z},
\end{equation}
$t(\textbf{a}_1) t(\textbf{a}_2) = t(\textbf{a}_2) t(\textbf{a}_1) = -t(\textbf{a}_1+\textbf{a}_2)$.
In this special case, there exist eigenstates $\left|n\textbf{k}\right>$ that simultaneously diagonalize $t(\textbf{a}_1)$, $t(\textbf{a}_2)$ and $H$, 
\begin{equation} \label{eq: def quasi-Bloch}
    t(\textbf{a}_1) \psi_{n,\textbf{k}}(\textbf{r}) = e^{-i\sigma\phi_1} e^{-i\textbf{k}\cdot\textbf{a}_1} \psi_{n,\textbf{k}}(\textbf{r}), \ \ \  
    t(\textbf{a}_2) \psi_{n,\textbf{k}}(\textbf{r}) = e^{-i\sigma\phi_2} e^{-i\textbf{k}\cdot\textbf{a}_2} \psi_{n,\textbf{k}}(\textbf{r}),
\end{equation}
where $\psi_{n,\textbf{k}}(\textbf{r})$ are the wavefunctions.
These eigenstates are labeled by the level index $n$ and a momentum $\textbf{k}$ defined by the phases of the eigenvalues of $t(\textbf{a}_{1,2})$. 
Because $t(\textbf{a})$ commutes with $a^\dagger a$, these states are tensor products of $\left|n\right>$ and $\left|\textbf{k}\right>$, the latter forming the quasi-Bloch representation for Landau level $n$.
The momentum of a normal Bloch wavefunction $\psi_\textbf{k}
(\textbf{r})=e^{i\textbf{k}\cdot\textbf{r}}u_\textbf{k}(r)$ is defined by its 
plane-wave envelope function.  Quasi-Bloch wavefunctions cannot be separated into an envelope and a periodic function, 
so we have the freedom to shift the momentum of all states by introducing 
the phases $\phi_{1,2}$ in all eigenvalues.
We use the convention $\phi_1 = \phi_2 = \pi$ such that $\left< \textbf{r}=0 \middle| \textbf{k}=0 \right> =0$.

The amplitudes of quasi-Bloch states $|\psi_{n,\textbf{k}}(\textbf{r})|^2$ are,
as in the ordinary Bloch state case, gauge-independent and periodic in the base lattice.
Note that for any lattice vector $\textbf{a}$, if we choose
the Landau gauge with vector potential 
\begin{equation}
    \textbf{A} = R_{-\theta(\textbf{a})}
    \begin{bmatrix}
    0 & \sigma B\\
    0 & 0
    \end{bmatrix}
    R_{\theta(\textbf{a})} \textbf{r},
\end{equation}
where $\theta(\textbf{a})$ is the angle from $+x$-axis to $\textbf{a}$,
then the magnetic translation operator $t(\textbf{a})$ is reduced to the normal translation operator $T(\textbf{a})$\footnote{In the definition, operator $R_{\theta}$ rotates two-component vectors counterclockwise by $\theta$.}.
For translation by the lattice vector $\textbf{a}=m\textbf{a}_1+n\textbf{a}_2$, 
\begin{equation}
    t(\textbf{a}) \psi_{n,\textbf{k}}(\textbf{r}) = \eta_{\textbf{a}}\, e^{-i\textbf{k}\cdot\textbf{a}} \psi_{n,\textbf{k}}(\textbf{r}),
\end{equation}
 where
 \begin{equation}
     \eta_{\textbf{a}} = \exp[i(m\sigma\phi_1+n\sigma\phi_2+mn\pi)] = \left\{
\begin{aligned}
    &+1,\ \ \text{$m$ and $n$ are even},\\
    &-1,\ \ \text{otherwise}.\\
\end{aligned}
     \right.
 \end{equation}
For our choice of $\phi_{1,2}$, the expression for $\eta$ 
is dependent on the lattice but independent of the choice of lattice generators $\textbf{a}_{1,2}$.
 
Consider a general magnetic translation on quasi-Bloch states
written in the form $t(-\sigma\hat{z}\times \textbf{k}l_B^2) \left|n\textbf{k}_1\right>$. 
Note that the displacement argument is a lattice vector when
\textbf{k} is a reciprocal lattice vector.
Using the commutator relation in Eq.~\ref{eq:translation commutator}, one can check that
the state produced by this operator acting on a Bloch state 
is another eigenstate of the lattice vector translation.
\begin{equation}
\begin{aligned}
    &t(\textbf{a}_1) \Big[ t(-\sigma\hat{z}\times \textbf{k}l_B^2) \psi_{n,\textbf{k}_1}(\textbf{r}) \Big]= -e^{-i(\textbf{k}+\textbf{k}_1)\cdot\textbf{a}_1}  \Big[ t(-\sigma\hat{z}\times \textbf{k}l_B^2) \psi_{n,\textbf{k}_1}(\textbf{r}) \Big], \\
    &t(\textbf{a}_2)  \Big[ t(-\sigma\hat{z}\times \textbf{k}l_B^2) \psi_{n,\textbf{k}_1}(\textbf{r}) \Big] = -e^{-i(\textbf{k}+\textbf{k}_1)\cdot\textbf{a}_2}  \Big[ t(-\sigma\hat{z}\times \textbf{k}l_B^2) \psi_{n,\textbf{k}_1}(\textbf{r}) \Big].\\
\end{aligned}
\end{equation}
One can deduce, by recognizing the eigenvalues, that $t(-\sigma\hat{z}\times \textbf{k}l_B^2) \left|n\textbf{k}_1\right>$ is, up to a global phase factor, the quasi-Bloch state $\left|n\textbf{k}_2\right>$ with momentum $\textbf{k}_2=\textbf{k}_1+\textbf{k}$. (Notice that $t(-\sigma\hat{z}\times \textbf{k}l_B^2) \left|n\textbf{k}_1\right>$ is normalized as $t(\textbf{a})$ is an unitary operator.)
We introduce the momentum shift operator as
\begin{equation} \label{eq: def tau(k)}
    \tau(\textbf{k}) = t(-\sigma\hat{z}\times \textbf{k}l_B^2) = e^{i\textbf{k}\cdot\textbf{C}},
\end{equation}
and apply the convention that fixes the global phase where
\begin{equation}\label{eq:phase convention}
    \left|\textbf{k}\right> = \tau(\textbf{k}) \left|\textbf{0}\right>.
\end{equation}

The multiplication of momentum shift operators $\tau(\textbf{k})$ gives an extra phase when switching the order, similar to magnetic translation operators.
\begin{equation}\label{eq:momentum commutator}
    \tau({\textbf{k}_1}) \tau({\textbf{k}_2}) = \tau({\textbf{k}_2}) \tau({\textbf{k}_2})e^{i\sigma(\textbf{k}_1\times\textbf{k}_2)_z/l_B^2} = \tau({\textbf{k}_1+\textbf{k}_2}) e^{i\frac{1}{2}\sigma(\textbf{k}_1\times\textbf{k}_2)_z/l_B^2}.
\end{equation}
If $\textbf{G}_1$ and $\textbf{G}_2$ form the reciprocal lattice such that $a_i \cdot G_j = 2\pi\delta_{ij}$,
\begin{equation}
    \textbf{G}_1 = -\hat{z} \times \textbf{a}_2 / l_B^2,\ \ \textbf{G}_2 = \hat{z} \times \textbf{a}_1 / l_B^2,\ \ \textbf{G}_1\times\textbf{G}_2 = -2\pi \hat{z} / l_B^2.
\end{equation}
In this lattice, $\tau({\textbf{G}_1}) \tau({\textbf{G}_2}) = \tau({\textbf{G}_2}) \tau({\textbf{G}_1}) = -\tau({\textbf{G}_1+\textbf{G}_2}) $.
Actually,
\begin{equation}
    \tau({\textbf{G}_1}) = t(-\sigma\textbf{a}_2),\ \ \tau({\textbf{G}_2}) = t(\sigma\textbf{a}_1).
\end{equation}
Shifting momentum by a general reciprocal lattice vector $\textbf{G} = m\textbf{G}_1 + n\textbf{G}_2$,
\begin{equation}
    \tau({\textbf{G}}) \psi_{n,\textbf{k}}(\textbf{r}) = \eta_{\textbf{G}} \,e^{i\sigma(\textbf{G}\times\textbf{k})_z l_B^2} \psi_{n,\textbf{k}}(\textbf{r}),
\end{equation}
 where
 \begin{equation}
     \eta_{\textbf{G}} = \left\{
\begin{aligned}
    &+1,\ \ \text{$m$ and $n$ are even},\\
    &-1,\ \ \text{otherwise}.\\
\end{aligned}
     \right.
 \end{equation}

Using the convention \ref{eq:phase convention} and the commuting rule \ref{eq:momentum commutator},
\begin{equation}
    \left|\textbf{k}_2\right> = \tau({\textbf{k}_2}) \tau({-\textbf{k}_1}) \left|\textbf{k}_1\right> = e^{-\frac{i}{2}\sigma (\textbf{k}_2\times\textbf{k}_1)_z l_B^2}\, \tau({\textbf{k}_2-\textbf{k}_1}) \left|\textbf{k}_1\right>.
\end{equation}
Specificly, if $\textbf{k}_2 - \textbf{k}_1$ is a reciprocal lattice vector,
\begin{equation} \label{eq:boost by G}
    \left|\textbf{k}+\textbf{G},\sigma\right> = e^{-\frac{i}{2}\sigma (\textbf{G}\times\textbf{k})_z l_B^2}\, \tau(\textbf{G}) \left|\textbf{k},\sigma\right>=\eta_{\textbf{G}} \,e^{\frac{i}{2} \sigma(\textbf{G}\times\textbf{k})_z l_B^2} \left|\textbf{k},\sigma\right>.
\end{equation}

For a general interaction, the mean-field calculations described in the main 
text require employ the following expression for $e^{-i\textbf{q}\cdot\textbf{r}}$ 
in the representation of quasi-Bloch states:
\begin{equation}
\begin{aligned}
        \left< n^\prime\textbf{k}^\prime \middle| e^{-i\textbf{q}\cdot\textbf{r}} \middle| n\textbf{k} \right> &= F^\sigma_{n^\prime,n}(-\textbf{q}) \left< \textbf{k}^\prime \middle| \tau({-\textbf{q}}) \middle| \textbf{k} \right>\\
        &= F^\sigma_{n^\prime,n}(-\textbf{q}) \left< 0 \middle| \tau(-\textbf{k}^\prime) \tau({-\textbf{q}}) \tau(\textbf{k}) \middle| 0 \right>\\
        &= F^\sigma_{n^\prime,n}(-\textbf{q}) e^{\frac{i}{2}\sigma (\textbf{k}^\prime\times\textbf{k})_z l_B^2} \sum_{\textbf{G}} \eta_{\textbf{G}} \,
        e^{ \frac{i}{2} \sigma \big( (\textbf{k}^\prime + \textbf{k})\times\textbf{G}\big)_{z} l_B^2} \delta_{\textbf{k}^\prime-\textbf{k}, -\textbf{q}+\textbf{G}}\\
        &= F^\sigma_{n^\prime,n}(-\textbf{q}) e^{\frac{i}{2}\sigma \big( \textbf{k}\times\textbf{k}^\prime + (\textbf{k}^\prime + \textbf{k})\times \textbf{q} \big)_z l_B^2} \sum_{\textbf{G}} \eta_{\textbf{G}} \,
        \delta_{\textbf{k}^\prime-\textbf{k}, -\textbf{q}+\textbf{G}},\\
        &= F^\sigma_{n^\prime,n}(-\textbf{q}) \sum_{\textbf{G}} 
        e^{\frac{i}{2}\sigma \big( \textbf{k}^\prime\times\textbf{k} + (\textbf{k}^\prime + \textbf{k})\times \textbf{G} \big)_z l_B^2}\eta_{\textbf{G}} \,
        \delta_{\textbf{k}^\prime-\textbf{k}, -\textbf{q}+\textbf{G}},\\
\end{aligned}
\end{equation}
where $\textbf{G}$ sums over all reciprocal lattice vectors, and the Landau level form factor
\begin{equation} 
\begin{aligned}
    &F^\sigma_{n_1,n_2}(q_x,q_y) 
    =\left\{
    \begin{aligned}
        \sqrt\frac{n_1!}{n_2!} \left(\frac{iq_x+\sigma q_y}{\sqrt{2}}l_B\right)^{n_2-n_1} e^{-\frac{q^2l_B^2}{4}} L_{n_1}^{(n_2-n_1)}&\left(\frac{q^2l_B^2}{2}\right),\,n_1\le n_2\\
        \sqrt\frac{n_2!}{n_1!} \left(\frac{iq_x-\sigma q_y}{\sqrt{2}}l_B\right)^{n_1-n_2} e^{-\frac{q^2l_B^2}{4}} L_{n_2}^{(n_1-n_2)}&\left(\frac{q^2l_B^2}{2}\right),\,n_1\ge n_2\\
    \end{aligned}
    \right.\\
    =\,&F^\sigma_{n_1,n_2}(q,\theta_\textbf{q}) = \sqrt\frac{n_<!}{n_>!} \left(\frac{iql_B}{\sqrt{2}}\right)^{n_>-n_<} e^{-i\sigma\theta_\textbf{q}(n_1-n_2)} e^{-\frac{q^2l_B^2}{4}} L_{n_<}^{(n_>-n_<)}\left(\frac{q^2l_B^2}{2}\right).
\end{aligned} 
\end{equation}
For comparison, we also list the matrix elements of the Fourier transform kernel in the representations of guiding center and angular momentum.
\begin{equation}
    \left< n^\prime X^\prime \middle| e^{-i\textbf{q}\cdot\textbf{r}} \middle| n X \right> = F^\sigma_{n^\prime,n}(-\textbf{q}) e^{-iq_x\frac{X+X^\prime}{2}} \delta_{X^\prime-X, -\sigma q_y l_B^2},
\end{equation}
and
\begin{equation}
    \left< n^\prime m^\prime \middle| e^{-i\textbf{q}\cdot\textbf{r}} \middle| n m \right> = F^\sigma_{n^\prime,n}(-\textbf{q}) F^\sigma_{m,m^\prime}(-\textbf{q}),
\end{equation}
where $X$ is the eigenvalue of $x$-direction guiding center operator $m$ is the eigenvalue of $b^\dagger b$.

Finally, we calculate the Berry connection $\textbf{A}_n(\textbf{k})$ and Chern number of the Landau level based on the convention in eq.~\ref{eq:phase convention}.
Consider an infinitesimal momentum step $d\textbf{k}$.
\begin{equation}
    e^{i\textbf{A}_n(\textbf{k})\cdot d\textbf{k}}=\left<n\ \textbf{k}+d\textbf{k} \middle| e^{id\textbf{k}\cdot \textbf{r}} \middle|n\,\textbf{k} \right>
    = e^{-\frac{i}{2}\sigma (\textbf{k} \times d\textbf{k})_z l_B^2}
    = e^{-\frac{i}{2}\sigma (\hat{z}\times\textbf{k})l_B^2 \cdot d\textbf{k} }
\end{equation}
Berry curvature therefore is $-\sigma l_B^2$, independent of $\textbf{k}$,
and the Chern number is $-\sigma$. This result is independent of the level index $n$.
 
\subsection{Wavefunctions}
In the lowest Landau level, the wavefunction can be constructed through the modified Weierstrass function in the symmetric gauge.
In symmetric gauge ($A_x = \sigma By/2$, $A_y = -\sigma Bx/2$) and lowest level case, for $\sigma=1$
\begin{equation}
    \psi_{\textbf{k}}^{+} (\textbf{r})= \mathcal{N} \tilde{\sigma}(z-z_\textbf{k}) \exp\left( \frac{z_\textbf{k}^* z}{2} - \frac{|z|^2+|z_\textbf{k}|^2}{4} \right),
\end{equation}
where $\tilde{\sigma}(z)$ is the modified Weierstrass function introduced in \cite{haldane2018modular} which reduces to the original Weierstrass function for square or triangular lattices, $z = (x+iy)/l_B$ and $z_\textbf{k} = \sigma(k_y-ik_x)l_B$, and $\mathcal{N}$ is the normalization symbol which implies different factors in different formula.
The $\sigma=-1$ wavefunction can be obtined by the time-reversal transformation $\psi_{\textbf{k}}^{-} (\textbf{r}) = \left[ \psi_{-\textbf{k}}^{+} (\textbf{r}) \right]^*$.
\begin{equation}
    \psi_{\textbf{k}}^{-} (\textbf{r}) = \mathcal{N} \tilde{\sigma}(z^*-z_\textbf{k}^*) \exp\left( \frac{z_\textbf{k} z^*}{2} - \frac{|z|^2+|z_\textbf{k}|^2}{4} \right).
\end{equation}
Applying $a^\dagger$ to these generates the wavefunctions of higher Landau levels.

The quasi-Bloch wavefunction can also be constructed from the guiding center representation $\left|X\right>$, using 
\begin{equation}
t(\textbf{a})\left|X\right> = \exp{\{-i\sigma a_y(X+a_x/2)/l_B^2\} \left|X+a_x\right>}    
\end{equation}
derived from the commutator in Eq.~\ref{eq:commutators}, where $\textbf{a}=(a_x,a_y)$.
The real and reciprocal lattice are generated by $\textbf{a}_1=(a_{1x}, a_{1y})$, $\textbf{a}_2=(a_{2x}, a_{2y})$, $\textbf{G}_1=(G_{1x}, G_{1y})$, and $\textbf{G}_2=(G_{2x}, G_{2y})$.
Define $\textbf{G}_\phi = \frac{\phi_1}{2\pi}\textbf{G}_1+\frac{\phi_2}{2\pi}\textbf{G}_2=(G_{\phi x}, G_{\phi y})$ and consider the following construction, where the level index is implicit,
\begin{equation}
\begin{aligned}
    \left| \textbf{k} \right> = \mathcal{N}\ \exp{\Big( i \, \sigma\alpha(\sigma\textbf{k},\textbf{G}_\phi) \Big) }\sum_{m,n = -\infty}^{+\infty} f_{mn}^\sigma\, & \exp{\Big( i ( k_x+ \sigma G_{\phi x} )\cdot X_{mn}(\sigma\textbf{k},\textbf{G}_\phi) \Big)} \Big| X_{mn}(\sigma\textbf{k},\textbf{G}_\phi) \Big>.
\end{aligned}
\end{equation}
By taking the coefficients
\begin{equation}
\begin{aligned}
    f_{mn}^\sigma &= \exp{\left( -\frac{i\sigma}{2l_B^2}\big(2mn a_{1x}a_{2y} + m^2 a_{1x}a_{1y} + n^2  a_{2x}a_{2y} \big)  \right)} \\
    &= \exp{\left( -\frac{i\sigma}{2l_B^2}\big(2mn a_{1y}a_{2x} + m^2 a_{1x}a_{1y} + n^2  a_{2x}a_{2y} \big)  \right)}
\end{aligned}
\end{equation}
together with the guiding centers
\begin{equation}
    X_{mn}(\sigma\textbf{k},\textbf{G}_\phi) = m a_{1x} + n a_{2x} + \sigma k_y l_B^2 + G_{\phi y} l_B^2,
\end{equation}
$\left| \textbf{k} \right>$ becomes an eigenstate of $t(\textbf{a}_{1,2})$ in Eq.~\ref{eq: def quasi-Bloch} with overall phase
\begin{equation}
    \alpha(\sigma\textbf{k},\textbf{G}_\phi) = \alpha(0,\textbf{G}_\phi) - \frac{1}{2} (\sigma k_x) (\sigma k_y) l_B^2 - G_{\phi x} (\sigma k_y) l_B^2
\end{equation}
in accord with the convention of Eq.~\ref{eq:phase convention}, assuming the nomalizing factor contains no $\textbf{k}$-dependent phase.
If $\alpha(0,\textbf{G}_\phi) = -G_{\phi x}G_{\phi y}l_B^2$,
\begin{equation}
\begin{aligned}
    \left| \textbf{k} \right> = \mathcal{N}\ e^{\frac{i\sigma}{2} k_x k_y l_B^2 }\sum_{m,n = -\infty}^{+\infty} f_{mn}^\sigma\, & e^{ i ( k_x+ \sigma G_{\phi x} ) (m a_{1x} + n a_{2x})} e^{ i k_x G_{\phi y} l_B^2}  \Big| X_{mn}(\sigma\textbf{k},\textbf{G}_\phi) \Big>.
\end{aligned}
\end{equation}

For the moir\'e triangular lattice in the main text, $\textbf{a}_1=(\sqrt{3}a/2, -a/2)$, $\textbf{a}_2=(0,a)$, $\textbf{G}_1=(G,0)$, and 
$\textbf{G}_2=(G/2, \sqrt{3}G/2)$, where $G = 4\pi/\sqrt{3}a$. 
As $\phi_1=\phi_2=\pi$, $\textbf{G}_\phi = (3G/4, \sqrt{3}G/4)$.
\begin{equation}
\begin{aligned}
    \left| \textbf{k} \right> = \mathcal{N}\ e^{ \frac{i\sigma}{2}k_xk_yl_B^2 }\sum_{m= -\infty}^{+\infty} (i\sigma)^{m(m-1)}\, & e^{ i k_x \left(m+\frac{1}{2}\right)\frac{\sqrt{3}a}{2}} \left| X=\left(m+\frac{1}{2}\right)\frac{\sqrt{3}a}{2} + \sigma k_y l_B^2 \right>.
\end{aligned}
\end{equation}

\section{Hartree-Fock Hamiltonian for Multiple Landau Levels} \label{apx: more Landau level}

Consider the moiré topological insulator with two valley components. 
Electrons in opposite valleys experience magnetic fields of the same magnitude but opposite 
direction. Below we replace the direction of the magnetic field $\sigma$
by the valley index $s$.

A general valley-conserving, valley-independent interaction written in the representation of quasi-Bloch states is
\begin{equation}
\begin{aligned}
        H_\text{int}= \frac{1}{2} \sum_{\alpha \beta \gamma \delta} v&_{\alpha \beta \gamma \delta} \,c^\dagger_{\alpha} c^\dagger_{\beta} c_{\gamma}  c_{\delta} \\
        =\frac{1}{2A} \sum_{s_1 s_2}& \sum_{n_1n_2n_3n_4} \sum_{\textbf{k}_1\textbf{k}_2\textbf{k}_3\textbf{k}_4} \int d^2\textbf{q} \, V(q) \\
        &\ \ \left< s_1n_1\textbf{k}_1 \middle| e^{-i\textbf{q}\cdot\textbf{r}} \middle| s_1 n_4\textbf{k}_4 \right>  \left< s_2n_2\textbf{k}_2 \middle| e^{i\textbf{q}\cdot\textbf{r}} \middle| s_2n_3\textbf{k}_3 \right> c^\dagger_{s_1n_1 \textbf{k}_1} c^\dagger_{s_2n_2 \textbf{k}_2} c_{s_2n_3 \textbf{k}_3}  c_{s_1n_4 \textbf{k}_4}.
\end{aligned}
\end{equation}
This interaction Hamiltonian can be treated using a 
Hartree-Fock approximation. When looking for a periodic state with the same 
periodicity as the quasi-Bloch states, we expect that each single-particle wavefunction in
the Slater determinant to have definite Bloch momentum $k$. 

The Hartree interaction is
\begin{equation}
\begin{aligned}
    \frac{1}{A} \sum_{ s n^\prime n \textbf{k}} \left(\sum_{s^\prime m^\prime m \textbf{p}} \sum_\textbf{G} V(\textbf{G}) F^s_{n^\prime n}(-\textbf{G}) F^{s^\prime}_{m^\prime m}(\textbf{G}) e^{i\big( (s \textbf{k} - s^\prime \textbf{p}) \times \textbf{G} \big)_z l_B^2} \left<c^\dagger_{s^\prime m^\prime \textbf{p}} c_{s^\prime m \textbf{p}}\right> \right) c^\dagger_{s n^\prime \textbf{k}} c_{s n \textbf{k}}.
\end{aligned}
\end{equation}
The intravalley Fock interaction is
\begin{equation}
\begin{aligned}
    - \frac{1}{A} \sum_{s n^\prime n \textbf{k}} \left(\sum_{ m^\prime m \textbf{p}} \sum_\textbf{G} V(\textbf{k}-\textbf{p}+\textbf{G}) F^s_{m^\prime n}(\textbf{p}-\textbf{k}-\textbf{G}) F^{s}_{n^\prime m}(\textbf{k}-\textbf{p}+\textbf{G})  \left<c^\dagger_{s m^\prime \textbf{p}} c_{s m \textbf{p}}\right> \right) c^\dagger_{s n^\prime \textbf{k}} c_{s n \textbf{k}},
\end{aligned}
\end{equation}
whereas the intervalley Fock interaction has an extra phase 
\begin{equation}
\begin{aligned}
    -\frac{1}{A} \sum_{s n^\prime n \textbf{k}} \left(\sum_{ m^\prime m \textbf{p}} \sum_\textbf{G} V(\textbf{k}-\textbf{p}+\textbf{G}) \right. F^s_{m^\prime n}&(\textbf{p}-\textbf{k}-\textbf{G}) F^{\Bar{s}}_{n^\prime m}(\textbf{k}-\textbf{p}+\textbf{G}) \\
    &  \times \  e^{is \big(\textbf{p}\times\textbf{k}+(\textbf{p}+\textbf{k})\times\textbf{G}\big)_z l_B^2} \left<c^\dagger_{s m^\prime \textbf{p}} c_{\Bar{s} m \textbf{p}}\right> \left. \vphantom{\sum_{ m^\prime m \textbf{p}}} \right) c^\dagger_{\Bar{s} n^\prime \textbf{k}} c_{s n \textbf{k}},
\end{aligned}
\end{equation}
that generally speaking results in inter-valley exchange that is weaker
than intra-valley exchange.

\section{Analysis of Two-Component Mean-Field Coherence}

\subsection{Berry Connection and Berry Curvature} \label{apx: Berry}

A well-known conclusion is that the Berry curvature of a two-band system is proportional
to the area enclosed on a two-level Bloch sphere per area in momentum space, but this is valid only when the basis bands have no Berry curvature, {\it e.g.} the case 
two sublattices in a tight-binding model.
In our problem, the basis bands are Landau levels.
The eigenstates of the Hartree-Fock renormalized lower band
\begin{equation}
    \left|\textbf{k} \right> = u_{\textbf{k}} \left|\textbf{k}\uparrow\right> + v_{\textbf{k}} \left|\textbf{k}\downarrow\right>,
\end{equation}
where $|u_\textbf{k}|^2 + |v_\textbf{k}|^2 = 1$.
Suppose we choose the global phases of different $\textbf{k}$ such that the coefficients $u_\textbf{k}$ and $v_\textbf{k}$ are differentiable functions of momentum.
Increasing $\textbf{k}$ by an infinitesimal $d\textbf{k}$, the basis Landau-level wavefunctions
\begin{equation}
    \left|\textbf{k}+d\textbf{k}\,s\right> = 
    e^{id\textbf{k}\cdot\textbf{r}}\Big( 1- i\textbf{A}_{\textbf{k}s}\cdot d\textbf{k} + O(|d\textbf{k}|^2) \Big) \left|\textbf{k}\, s\right> , s=\uparrow\text{or}\downarrow,
\end{equation}
and the eigenstates
\begin{equation}
\begin{aligned}
    \left|\textbf{k}+d\textbf{k} \right>\,& =  u_{\textbf{k}+d\textbf{k}} \left|\textbf{k}+d\textbf{k}\uparrow\right> + v_{\textbf{k}+d\textbf{k}} \left|\textbf{k}+d\textbf{k}\downarrow\right>\\
    &= e^{id\textbf{k}\cdot\textbf{r}} \Big( 
    \left|\textbf{k} \right> + 
    \partial_\textbf{k}u_\textbf{k} \cdot d\textbf{k} \left|\textbf{k}\uparrow\right> +
    \partial_\textbf{k}v_\textbf{k} \cdot d\textbf{k} \left|\textbf{k}\downarrow\right> -
     iA_{\textbf{k}\uparrow}\cdot d\textbf{k}\, u_\textbf{k} \left|\textbf{k}\uparrow\right> -
    iA_{\textbf{k}\downarrow}\cdot d\textbf{k}\, v_\textbf{k} \left|\textbf{k}\downarrow\right> +
    O(|d\textbf{k}|^2) \Big)
\end{aligned}
\end{equation}
By definition, the Berry connection satisfies
\begin{equation}
\begin{aligned}
    1+i A_\textbf{k}\cdot d\textbf{k} = \left< \textbf{k}+d\textbf{k} \middle| e^{id\textbf{k}\cdot\textbf{r}} \middle| \textbf{k} \right>
    = & 1+ u_\textbf{k} \partial_\textbf{k} u^*_\textbf{k} \cdot d\textbf{k} + 
    v_\textbf{k} \partial_\textbf{k} v^*_\textbf{k} \cdot d\textbf{k} 
    +i A_{\textbf{k}\uparrow}\cdot d\textbf{k}     |u_\textbf{k}|^2 +i A_{\textbf{k}\downarrow}\cdot d\textbf{k} |v_\textbf{k}|^2 \\
    A_\textbf{k} & = -iu_\textbf{k} \partial_\textbf{k} u^*_\textbf{k} -i v_\textbf{k} \partial_\textbf{k} v^*_\textbf{k} + |u_\textbf{k}|^2 A_{\textbf{k}\uparrow} + |v_\textbf{k}|^2 A_{\textbf{k}\downarrow}.
\end{aligned}
\end{equation}
There are contributions from the Berry connections of the 
basis bands to the full Berry connection and, therefore, the Berry curvature.
Similar results can be derived for multiple-component systems, including the mixing of multiple Landau levels, charge density wave/Wigner crystals, and the Hofstadter problem.

\subsection{Inter-Band Coherence Phase Winding} \label{apx:phase winding number}

For the Landau-level problem that we research in this paper, the two singularities in $\phi_\textbf{k}$ are a direct consequence of Eq.~\ref{eq:phi plus G} derived from Eq.~\ref{eq:boost by G}.
This appendix extends this conclusion to the phase winding in the coherence 
between two general bands.
In this case, the winding number is exactly the difference in the Chern numbers of the two bands, independent of the details of their quantum geometry.

We have basis-band Bloch wavefunctions $\{\left| \textbf{k}, i\right>\}$, where $i$=1 or 2 denotes the two band.
Usually, we care only about the momentum reduced into the Brillouin zone, as the states differing by reciprocal lattice vectors are the sam states.
However, by tuning the overall phase in the wavefunctions of different momentum, $\psi_{\textbf{k}i}(\textbf{r}) = \left< \textbf{r} \middle| \textbf{k}, i \right>$ can be a smooth function of both $\textbf{r}$ and $\textbf{k}$, where $\textbf{k}$ is smoothly extended outside the boundary of the Brillouin zone or reciprocal unit cell.
In this case, the wavefunction acquires an extra phase $\alpha$ when boosted by a reciprocal lattice vector, {\it i.e.}, 
\begin{equation}
    \psi_{\textbf{k}+\textbf{G},i}(\textbf{r}) = e^{i\alpha_i(\textbf{k},\textbf{G})} \psi_{\textbf{k}i}(\textbf{r})\,.
\end{equation}
If the Berry connections of the basis bands are $A_i(\textbf{k})$, 
\begin{equation}
    A_i(\textbf{k})-A_i(\textbf{k}+\textbf{G}) = {\bf \nabla}_\textbf{k} \alpha_i(\textbf{k},\textbf{G})\,.
\end{equation}
The Chern numbers $C_i$ are calculated by integrating the Berry connections along the boundary of the reciprocal unit cell, $0 \rightarrow \textbf{G}_1 \rightarrow \textbf{G}_1+\textbf{G}_2 \rightarrow \textbf{G}_2 \rightarrow 0$.
The basic vectors $\textbf{G}_1$ and $\textbf{G}_2$ are chosen such that this contour loop is in counterclockwise direction.
\begin{equation}
\begin{aligned}
    2\pi C_i &= \left( \int_0^{\textbf{G}_1} + \int_{\textbf{G}_1}^{\textbf{G}_1+\textbf{G}_2} + \int_{\textbf{G}_1+\textbf{G}_2}^{\textbf{G}_2} + \int_{\textbf{G}_2}^0 \right) \textbf{A}_i(\textbf{k}) \cdot d\textbf{k}\\
    &=\int_0^{\textbf{G}_1} \left[ \textbf{A}_i(\textbf{k})-\textbf{A}_i(\textbf{k}+\textbf{G}_2)\right] \cdot d\textbf{k} + \int_0^{\textbf{G}_2} \left[ \textbf{A}_i(\textbf{k}+\textbf{G}_1)-\textbf{A}_i(\textbf{k})\right] \cdot d\textbf{k}\\
    &=\int_0^{\textbf{G}_1} {\bf \nabla}_\textbf{k}\alpha_i(\textbf{k},\textbf{G}_2) \cdot d\textbf{k} - \int_0^{\textbf{G}_2} {\bf \nabla}_\textbf{k}\alpha_i(\textbf{k},\textbf{G}_1) \cdot d\textbf{k}\,.\\
\end{aligned}
\end{equation}

When we find the interband coherent phase in the mean-field approximation, the eigenstates are
\begin{equation}
    \left|\textbf{k} \right> = u_{\textbf{k}} \left|\textbf{k},1\right> + v_{\textbf{k}} \left|\textbf{k},2\right>\ \ \sim \ \ \left|\textbf{k} + \textbf{G} \right> = u_{\textbf{k} + \textbf{G}} \left|\textbf{k} + \textbf{G},1\right> + v_{\textbf{k} + \textbf{G}} \left|\textbf{k} + \textbf{G},2\right>,
\end{equation}
where $|u_\textbf{k}|^2 + |v_\textbf{k}|^2 = 1$, and the coherence phase, the azimuthal angle of the Bloch sphere, is defined by 
\begin{equation}
    \phi_\textbf{k} = \arg(v_\textbf{k}/u_\textbf{k})\,.
\end{equation}
Because $\left|\textbf{k} \right>$ and $\left|\textbf{k} + \textbf{G} \right>$ only differ by an overall phase, 
\begin{equation}
\begin{aligned}
    \left|\textbf{k}+\textbf{G}\right> &= u_{\textbf{k}+\textbf{G}} \left|\textbf{k}+\textbf{G},1\right> + v_{\textbf{k}+\textbf{G}} \left|\textbf{k}+\textbf{G},2\right> \\
    &= u_{\textbf{k}+\textbf{G}} e^{i\alpha_1(\textbf{k},\textbf{G})} \left|\textbf{k},1\right> + v_{\textbf{k}+\textbf{G}} e^{i\alpha_2(\textbf{k},\textbf{G})} \left|\textbf{k},2\right>\\
    &= e^{i\alpha(\textbf{k},\textbf{G})} \Big( u_{\textbf{k}} \left|\textbf{k},+\right> + v_{\textbf{k}}  \left|\textbf{k},2\right> \Big) = e^{i\alpha(\textbf{k},\textbf{G})}\left|\textbf{k} \right>\,.\\
\end{aligned}
\end{equation}
To match the phases of the two sides of the third equation,
\begin{equation}
    \phi_{\textbf{k}+\textbf{G}} = \phi_\textbf{k} + \alpha_1(\textbf{k},\textbf{G}) - \alpha_2(\textbf{k},\textbf{G})\,.
\end{equation}

Now consider the winding number of $\phi_\textbf{k}$ also along the reciprocal unit cell.
\begin{equation}
\begin{aligned}
    2\pi W &= \left( \int_0^{\textbf{G}_1} + \int_{\textbf{G}_1}^{\textbf{G}_1+\textbf{G}_2} + \int_{\textbf{G}_1+\textbf{G}_2}^{\textbf{G}_2} + \int_{\textbf{G}_2}^0 \right) {\bf \nabla}_\textbf{k}\phi_\textbf{k} \cdot d\textbf{k}\\
    &=\int_0^{\textbf{G}_1} \left[ {\bf \nabla}_\textbf{k}\phi_\textbf{k} - {\bf \nabla}_\textbf{k}\phi_{\textbf{k}+\textbf{G}_2}\right] \cdot d\textbf{k} + \int_0^{\textbf{G}_2} \left[ {\bf \nabla}_\textbf{k}\phi_{\textbf{k}+\textbf{G}_1} - {\bf \nabla}_\textbf{k}\phi_\textbf{k}\right] \cdot d\textbf{k}\\
    &=\int_0^{\textbf{G}_1} [{\bf \nabla}_\textbf{k}\alpha_2(\textbf{k},\textbf{G}_2) - {\bf \nabla}_\textbf{k}\alpha_1(\textbf{k},\textbf{G}_2)] \cdot d\textbf{k} - \int_0^{\textbf{G}_2} [{\bf \nabla}_\textbf{k}\alpha_2(\textbf{k},\textbf{G}_1) - {\bf \nabla}_\textbf{k}\alpha_1(\textbf{k},\textbf{G}_2)] \cdot d\textbf{k}\\
    &= 2\pi (C_2 - C_1).
\end{aligned}
\end{equation}
With this, we have proved the result promised at the beginning of this appendix.

\newpage
\twocolumngrid

\section{Crystal Symmetries of Landau-level Intervalley Coherent States} \label{apx:symmetry}

\subsection{Various Symmetries} \label{apx:crystal symmetry}

With the belief that states with certain crystal symmetries have at least locally minimum energies, we compute Hartree-Fock IVC solutions with all the possible crystal symmetries and compare their energies, as described in the main text.
In this appendix, we present the requirements for the Hartree-Fock density matrices (or $\theta_\textbf{k}$ and $\phi_\textbf{k}$) to exhibit a certain symmetry. 
By imposing these symmetries in the Hartree-Fock calculation, we can find the desired solutions.
As we are looking for the ground state, we will limit our search to the states with only two Dirac points because more Dirac points usually increase the energy.

Because we are working with the intervalley coherent phase with U(1)$_v$ symmetry {\it spontaneously} broken, all the symmetry transforms are allowed to be composed of a phase addition.\footnote{For a spontaneous-symmetry-breaking phase, the ground states span an ``ground subspace''. This ``ground subspace'' is invariant under the broken symmetry transform. In IVC, this is the global phase addition. If the Hartree-Fock state is conserved by a symmetry transform composed of a phase addition, then the subspace is conserved by the symmetry transform.}
For discrete symmetries, the possible phase additions are also discrete, on which will be elaborated below.

\subsubsection{Unit-Cell Lattice Translation Symmetry, Inversion and Time-Reversal Symmetries}

We first investigate the non-crystal symmetries that the solutions frequently exhibit. 
The first one is the translation symmetry related to the magnetic unit cell that we employed to construct the quasi-Bloch representation.
Due to Eq.~\ref{eq:boost by G}, 
\begin{equation}
\begin{aligned}
    &\phi_{\textbf{k}+\textbf{G}} = \phi_\textbf{k} +  (\textbf{G}\times\textbf{k})_z\ l_B^2,\\ 
    &\theta_{\textbf{k}+\textbf{G}} = \theta_\textbf{k},
\end{aligned}
\end{equation}
where $\textbf{G}$ is a reciprocal lattice vector.
This requires at least two phase singularities in the Brillouin zone which becomes two Dirac points.
As we are looking for the ground state, we will limit our search to the states with only two Dirac points because more Dirac points typically increase the energy.

The inversion (P) requires
\begin{equation}
\begin{aligned}
    &\phi_{-\textbf{k}} = \phi_\textbf{k} + n_1\pi,\\
    &\theta_{-\textbf{k}} = \theta_\textbf{k},
\end{aligned}
\end{equation}
and the time-reversal symmetry
\begin{equation}
\begin{aligned}
    &\phi_{-\textbf{k}} = \phi_\textbf{k} + n_1\pi,\\
    &\theta_{-\textbf{k}} = \pi - \theta_\textbf{k}.
\end{aligned}
\end{equation}
They differ in the transform of $\theta_\textbf{k}$ that controls how the Dirac points are polarized when a gap opens and the Chern number of gapped bands.
Integer $n_1$ in the transform can be 0 or 1, controlling the composed phase addition.
For each momentum $\textbf{X} = -\textbf{G}/2$ where $\textbf{G}$ is a reciprocal lattice vector,
\begin{equation} \label{eq: PT sym phase}
\begin{aligned}
        \phi_{\textbf{X}} + n_1 \pi = \phi_{-\textbf{X}} &= \phi_{\textbf{X}+\textbf{G}} \\
        &= \phi_{\textbf{X}} +  (\textbf{G}\times\textbf{X})_z\ l_B^2 = \phi_{\textbf{X}}.
\end{aligned}
\end{equation}
If $n_1 = 1$, $\textbf{X}$ is a phase singularity.
In the first Brillouin zone of moir\'e triangular lattice, possible $\textbf{X} = -\textbf{G}/2$ are $\gamma$($\textbf{k}=0$) and the three middle points (M) of the Brillouin zone edges -- at least four phase singularities in total.
To avoid such solutions, we will only consider $n_1=0$.

\subsubsection{C$_{3z}$ Rotational Symmetry}

The moir\'e lattice has three-fold rotational symmetry C$_{3z}$, allowing the IVC solution to be C$_{3z}$ symmetric.
The C$_{3z}$ symmetry requires
\begin{equation}
\begin{aligned}
&\phi_{R(120^\circ)\textbf{k}}=\phi_\textbf{k}+ n_2\,2\pi/3,\\
&\theta_{R(120^\circ)\textbf{k}} = \theta_\textbf{k},
\end{aligned}
\end{equation}
where $R(120^\circ)$ is the rotation operator for 2D vectors, and the integer $n_2$ can be 0, 1 or 2.
Consider the reciprocal lattice generated by $\textbf{G}_1 = (G,0)$ 
and $\textbf{G}_2 = (G/2, \sqrt3 G/2)$.
Similar to Eq.~\ref{eq: PT sym phase}, we consider the high symmetry points $\gamma$, $\kappa$, and $\kappa^\prime$, where $\kappa^{(\prime)}$ are the two corners of the magnetic Brillouin zone.
\begin{equation}
\begin{aligned}
    &\gamma=0, &&R(120^\circ)\gamma = \gamma;\\
    &\kappa=(\textbf{G}_1+\textbf{G}_2)/3, &&R(120^\circ)\kappa = \kappa-\textbf{G}_1;\\
    &\kappa'=2(\textbf{G}_1+\textbf{G}_2)/3, && R(120^\circ)\kappa' = \kappa'-2\textbf{G}_1.\\
\end{aligned}
\end{equation}
The coherence phases at these points under the rotational transform are
\begin{equation}
\begin{aligned}
    &\phi_\gamma+ n_2\, 2\pi/3 = \phi_\gamma,\\
    &\phi_\kappa+ n_2\, 2\pi/3 = \phi_{\kappa-\textbf{G}_1} \\
    &\ \ \ \ \ \ \ \ \ \ \ \ \ \ = \phi_\kappa - (\textbf{G}_1\times \kappa)_z\,l_B^2 = \phi_\kappa - 2\pi/3,\\
    &\phi_{\kappa'}+ n_2\, 2\pi/3 = \phi_{\kappa^\prime-2\textbf{G}_1} \\
    &\ \ \ \ \ \ \ \ \ \ \ \ \ \ = \phi_{\kappa^\prime} - (2\textbf{G}_1\times \kappa')_z\,l_B^2 = \phi_{\kappa'} - 8\pi/3.\\
\end{aligned}
\end{equation}
If $n_2=0$, $\kappa$ and $\kappa$' are the two singularities while $\gamma$ is not.
If $n_2=2$, $\gamma$ is a singularity while $\kappa$ and $\kappa$' are not.
Due to the P/T symmetry, if $\textbf{X}$ is a singularity, $-\textbf{X}$ also is, so $\gamma$ singularity has winding number 2 at least.
With these two symmetries, we can find self-consistent solutions, labeled by C$_{3z}$ and C$_{3z}+4\pi/3$ as listed in Table \ref{tab:crystal symmetry}.
On the other hand, if $n_2=1$, $\kappa$, $\kappa$' and $\gamma$ are all singularities.
However, because their winding numbers cannot add up to 2, other singularities must appear. 
With the C$_{3z}$ and P/T symmetries, the number of them must be a multiple of 6.  
Consequently, the number of singularities has to be at least 9.

\begin{figure}
    \label{fig:Abrikosov}
    \centering
    \includegraphics[width=\linewidth]{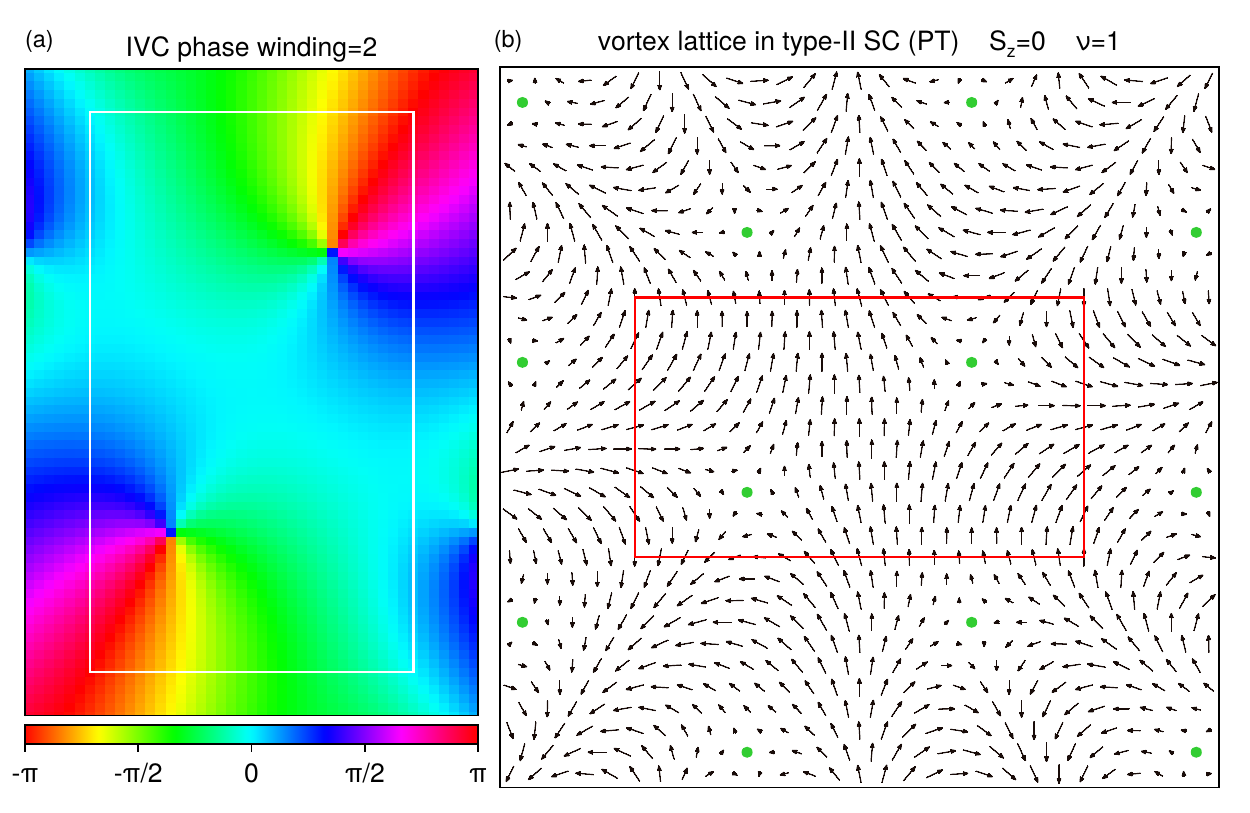}
    \caption{The IVC problem at $\lambda=0$ is equivalent to the superconductivity in the two-spin Landau level system. In the latter problem, the ground state is the Abrikosov's triangular vortex lattice. This figure shows the mean-field solution of IVC state with the same triangular lattice, which should be interpreted as a half-lattice translation symmetry in a rectangular lattice (see Appendix \ref{apx:SC VL}). (a) $\phi_\textbf{k}$ field in reciprocal space has two singularities. The white rectangle denotes the Brillouin zone. (b) order parameter shows vortex lattice pattern in real space. The vortex centers, marked by green dots, form a triangular lattice. The red rectangle denotes the unit cell.  }
\end{figure}

\subsubsection{Half Unit-Cell Lattice Translation Symmetry}
We know that the IVC states, using the Landau level wavefunction, are vortex lattices in real space.
In every unit cell, two vortices form the two sublattices. 
If the displacement between two sublattices is half of a lattice vector, this bipartite lattice becomes a simple lattice with half of the unit cell area.
In this case, the system have the mid-lattice translation symmetry, and the solution is invaritant under $t(\textbf{a}_0/2)$ or, with Eq.~\ref{eq: def tau(k)}, $\tau({\textbf{G}_0/2})$, where $\textbf{G}_0$ is a reciprocal lattice vector such that $\textbf{G}_0/2$ is not.
This requires
\begin{equation} \label{eq:half translation}
\begin{aligned}
    &\phi_{\textbf{k}+\textbf{G}_0/2} = \phi_\textbf{k} +  (\textbf{G}_0/2\times\textbf{k})_z\,l_B^2 + n_3 \pi,\\
    &\theta_{\textbf{k}+\textbf{G}_0/2} = \theta_\textbf{k},
\end{aligned}
\end{equation}
where $n_3$ = 0 or 1.
We notice that the $n_3=0$ case can be reduced to $n_3=1$ by replacing $\textbf{G}_0$ with $\textbf{G}_0+ 2\textbf{G}_0^\prime$, where $\textbf{G}_0^\prime$ is another reciprocal lattice vector such that $(\textbf{G}_0^\prime \times\textbf{G}_0)_zl_B^2=2\pi$.
Consequently, we will only focus on the latter case.

With the P/T symmetry,
\begin{equation} 
\begin{aligned}
        \phi_{-\textbf{G}_0/4}&= \phi_{\textbf{G}_0/4}\\
        &= \phi_{-\textbf{G}_0/4} +  [\textbf{G}_0/2\times(-\textbf{G}_0/4)]_zl_B^2 +\pi \\
        &= \phi_{-\textbf{G}_0/4} +\pi.
\end{aligned}
\end{equation}
So, the two singularities are located at $\pm\textbf{G}_0/4$.
For the moiré lattice, solutions of this type must break C$_{3z}$ rotational symmetry, ending up with three equivalent solutions up to $2\pi/3$ rotation.
The energy is also listed in Table \ref{tab:crystal symmetry} in the main text.

\subsection{Triangular vortex lattice in type-II superconductor (Lowest Landau level)} \label{apx:SC VL}

Superconductivity in two-dimensional Landau levels was studied extensively in previous works like \cite{ tesanovic1989new,akera1991vortexlattice},
where the two (spin) components are in the same magnetic field.
This was a natural extension of type-II superconductors in the strong field limit,
and the results agreed with the Ginzburg-Landau effective theory that the ground state should be a triangular vortex lattice.
In the main text, we argue that this problem is equivalent to our moir\'e topological insulator problem with the Landau-level idealization at $\lambda=0$.
Here, we present the method of calculating the triangular vortex lattice solution within our Hartree-Fock setup.

We first analyze the symmetry of this state.
It is defined in the magnetic lattice generated by 
\begin{equation}
\begin{aligned}
    &\textbf{a}_1=( \sqrt{3}a,0),\ \textbf{a}_2= ( 0,a), \, a=\sqrt{\frac{A_\Phi}{\sqrt{3}}},\\
    &\textbf{G}_1=(G,0),\ \textbf{a}_2=(0,\sqrt{3}G),\, G=\frac{2\pi}{\sqrt{3}a},
\end{aligned}
\end{equation}
with unbroken P and T symmetry, and is invariant under half-lattice translation in Eq.~\ref{eq:half translation} ($n_3$=1) with $\textbf{G}_0=\textbf{G}_1+\textbf{G}_2$.
We calculate the Hartree-Fock solution with these symmetries at $\lambda=0$, as shown in Fig. ~\ref{fig:Abrikosov}.
Panel (a) depicts $\phi_\textbf{k}$ in reciprocal space, where the white rectangle denotes the Brillouin zone.
The two phase singularities are removed by $\textbf{G}_0/2$.
Panel (b) shows the real-space order-parameter distribution with the vortex lattice pattern. 
The red rectangle denotes real-space unit cell, and the green dots denote the vortex centers that form a triangular lattice.

\onecolumngrid

\section{Time-dependent Hartree-Fock theory}
Time-dependent Hartree-Fock (TDHF) theory calculates the bosonic excitations of a mean-field theory under the Hartree-Fock approximation, given the mean-field ground state.
Here, we briefly formulate the TDHF method and apply it to the intervalley coherent states.
Using the Landau-level quasi-Bloch wavefunctions, we calculate the coefficients of the TDHF equations and solve them numerically.
We show that the breaking of electron number conservation of each valley/spin, {\it i.e.}, the U(1)$_z$ symmetry, leads to the spin-flipping magnon mode that satisfies the Goldstone theorem.

\subsection{Linear Response of The Mean-Field Theory}

We start with the linear response theory of a non-interacting system.
If the Hamiltonian is perturbed with the term $\hat{B}\,\epsilon e^{i\omega t}$, the expectation of the observable $\hat{A}$ is varied by an amount linear to $\epsilon e^{i\omega t}$ when $\epsilon\rightarrow0$.
The coefficient of this response is the correlation function $\chi^f_{AB}(\omega)$ whose Fourier transform 
\begin{equation}
    \int \frac{d\omega}{2\pi} e^{-i\omega t} \chi^f_{AB}(\omega) = -i \Theta(t)\left< \left[ \hat{A}(t), \hat{B}(0) \right]\right>,
\end{equation}
where $\left< \dots \right>=\text{Tr}(\rho_0 \dots)$ is the expectation evaluated with the unperturbed density matrix $\rho_0$.
Taking the time derivative on both sides,
\begin{equation}
\begin{aligned}
    \int \frac{d\omega}{2\pi} e^{-i\omega t} (-i\omega) \chi^f_{AB}(\omega) &= -i \Theta(t)\left< \left[ i\left[H,\hat{A}(t)\right], \hat{B}(0) \right]\right> -i \delta(t) \left< \left[ \hat{A}(0), \hat{B}(0) \right]\right>.
\end{aligned}
\end{equation}

We focus on the density-density responses
where both $\hat{A}$ and $\hat{B}$ are density operators in the form of $c^\dagger_\alpha c_\beta$.
The Greek letters denote single-particle states in a complete basis of the Hilbert space, and the density operators are labeled 
by pairs of states, which we call the excitation indices.
We employ the energy eigenstate representation so that the Hamiltonian $H = \sum_\alpha E_\alpha c^\dagger_\alpha c_\alpha$.
The commutator of two density operators
\begin{equation}
    C_{\alpha\beta,\gamma\delta} = \left< [c^\dagger_\alpha c_\beta, c^\dagger_\gamma c_\delta] \right> = \delta_{\alpha\delta} \left< c^\dagger_\gamma c_\beta \right> - \delta_{\beta\gamma} \left< c^\dagger_\alpha c_\delta \right>.
\end{equation}
The commutator of a density operator with $H$ 
\begin{equation}
    [H, c^\dagger_\alpha c_\beta] = -(E_\alpha-E_\beta)c^\dagger_\alpha c_\beta = -E_{\alpha\beta,\gamma\delta}\,c^\dagger_\gamma c_\delta
\end{equation}
appears as a linear transform on the density operators defined by a diagonal matrix of excitation energy
\begin{equation}
    E_{\alpha\beta,\gamma\delta} = (E_\alpha-E_\beta) \delta_{\alpha\beta,\gamma\delta}.
\end{equation}
$\chi^f(\omega)$, $C$, and $E$ are all matrices in excitation index; the previous equation is written with matrix multiplication as
\begin{equation}
\begin{aligned}
    (-i\omega) \chi^f(\omega) =& -iE\chi^f(\omega) -iC\\
    (\omega I - E)&\chi^f(\omega)=C,\\
\end{aligned}
\end{equation}
or 
\begin{equation}
    \sum_{\mu\nu}\Big(\omega \delta_{\alpha\beta,\mu\nu} - E_{\alpha\beta,\mu\nu}\Big)\chi^f_{\mu\nu,\gamma\delta}(\omega)=C_{\alpha\beta,\gamma\delta}.
\end{equation}

The theory above, albeit the standard formulation for linear response theory, is not directly applicable to a mean-field theory. 
Instead, the $\chi^f$ is understood as the correlation function for a fictitious non-interacting system whose Hamiltonian happens to equal the mean-field Hamiltonian at the HF ground state.
The difference lies in the mean-field term in the Hamiltonian that depends on the density matrix.
When a perturbation is introduced in a mean-field theory, the density matrix and the mean field are changed.
Equivalently, we spontaneously introduce two perturbations, the original one and the variation of the mean field.
The second part is linear to the variation of the density matrix and is calculated by the density response to the original perturbation.
The true response is the "fictitious" response to both perturbations.

Consider a density perturbation $c^\dagger_\gamma c_\delta \,\epsilon e^{i\omega t}$, the density matrix variation 
\begin{equation}
    \Delta \left< c^\dagger_\mu c_\nu\right> = \chi_{\mu\nu,\gamma\delta}(\omega) \, \epsilon e^{i\omega t}
\end{equation}
and the mean field variation
\begin{equation}
    \Delta H_\text{MF} = H_\text{MF}\big[\Delta \left< c^\dagger_\mu c_\nu \right>\big] = V_{\tau\sigma,\mu\nu} 
    \, \Delta\left< c^\dagger_\mu c_\nu\right> \, c^\dagger_\tau c_\sigma = 
    V_{\tau\sigma,\mu\nu} \, \chi_{\mu\nu,\gamma\delta}(\omega)\, c^\dagger_\tau c_\sigma \,\epsilon e^{i\omega t},
\end{equation}
where the interaction matrix $V$ is the linear coefficient of the mean field to the density matrix.
In the Hartree-Fock approximation, given the interaction $H_{int}=\frac{1}{2}v_{\alpha\beta\gamma\delta} c^\dagger_\alpha c^\dagger_\beta c_\gamma c_\delta$,
\begin{equation} \label{eq: matrix V}
    V_{\tau\sigma,\mu\nu} = v_{\tau\mu\nu\sigma} - v_{\mu\tau\nu\sigma}.
\end{equation}
The true correlation function
\begin{equation}
    \chi_{\alpha\beta,\gamma\delta}(\omega) = \chi^f_{\alpha\beta,\gamma\delta}(\omega) + \chi^f_{\alpha\beta,\tau\sigma}(\omega) \, V_{\tau\sigma,\mu\nu} \, \chi_{\mu\nu,\gamma\delta}(\omega)
\end{equation}
Multiplying $\omega I -E$ on both sides,
\begin{equation}\label{eq:TDHF 1}
\begin{aligned}
    (\omega I-E)\chi(\omega) = (\omega I -E)&\chi^f(\omega) + (\omega I-E)\chi^f(\omega)\cdot V\cdot\chi(\omega)\\
    (\omega I-E)\chi(\omega) &= C + CV\cdot\chi(\omega)\\
    \big[\omega I -(E+&CV)\big]\chi(\omega) = C\\
\end{aligned}
\end{equation}
or
\begin{equation}
    \sum_{\mu\nu}\left[\omega \delta_{\alpha\beta,\mu\nu} - \left( E_{\alpha\beta,\mu\nu} + \sum_{\tau\sigma} C_{\alpha\beta,\tau\sigma} V_{\tau\sigma,\mu\nu} \right) \right]\chi_{\mu\nu,\gamma\delta}(\omega)=C_{\alpha\beta,\gamma\delta}.
\end{equation}
To calculate the retarded response, we replace $\omega$ with $\omega + i\eta$.
The bosonic excitations of the ground state correspond to higher energy eigenstates. 
The excitation energies are the frequencies where the correlation function diverges and are the eigenvalues of the matrix $E+CV$.
Although it is not hermitian, its eigenvalues are all real and are in pairs of positive and negative numbers with the same absolute values, 
corresponding to the excitations and deexcitations that we will discuss later.

Matrix $V$ is usually not calculated in the energy eigenbasis. 
To perform the representation transform, we notice that the creation (annihilation) operators are covariant (contravariant).
We specify all the covariant and contravariant indices in the previous equation.
\begin{equation}
    \sum_{\mu\nu}\left[\omega \delta_\alpha^\mu  \delta_\nu^\beta  - \left( E_\alpha{}^\beta{}_,{}^\mu{}_\nu + \sum_{\tau\sigma} C_\alpha{}^\beta{}_{,\tau}{}^\sigma V^\tau{}_{\sigma,}{}^\mu{}_\nu \right) \right] \chi_\mu{}^\nu{}_{,\gamma}{}^\delta (\omega)=C_\alpha{}^\beta{}_{,\gamma}{}^\delta\,.
\end{equation}

\subsection{Application to Intervalley Coherence}
In the valley-ordered states of moir\'e topological insulators, the energy eigenbasis $\alpha$ consists of momentum $\textbf{k}$ and band index $c/v$.
Two simplifications are made for further calculations.
First, commutation matrix $C_{\alpha\beta,\gamma\delta} = \delta_{\alpha\delta} \delta_{\gamma\beta} \big[ n_F(\beta) - n_F(\alpha) \big]$. 
Therefore, at zero temperature, we only need to consider correlations between density operators across band indices, namely, the excitations $c^\dagger_{ c} c_{ v}$ and deexcitations $c^\dagger_{v} c_{c}$.
Second, because of the momentum conservation of the interaction $V$, the momentum transfer of the excitation and deexcitation must be opposite.
This transferred momentum is the momentum of the bosonic excitation.
All the correlation functions that fail to satisfy the two requirements vanish.

The TDHF matrix is decomposed by the excitation momentum $\textbf{q}$.
For each $\textbf{q}$, we formulate matrices in the 2-by-2 form of blocks, denoting the excitation and deexcitation degree of freedom, 
and each block is a matrix in the momentum degree of freedom.
We distinguish the momentum index $\textbf{k}$ for momentum transfer $\textbf{q}$ and index $\textbf{p}$ for $-\textbf{q}$.
The valance band momenta are $\textbf{k}/\textbf{p}$ and the conduction band momenta are $\textbf{k}/\textbf{p} \pm \textbf{q}$.
To be specific, index $\textbf{k}$ denotes $(\textbf{k}+\textbf{q}\,c,\ \textbf{k}\,v)$ in excitations and $(\textbf{k}\,v,\ \textbf{k}-\textbf{q}\,c)$ in deexcitations, 
and index $\textbf{p}$ denotes $(\textbf{p}-\textbf{q}\,c,\ \textbf{p}\,v)$ in excitations and $(\textbf{p}\,v,\ \textbf{p}+\textbf{q}\,c)$ in deexcitations.
In such indexing rules, the TDHF equation, eq.~\ref{eq:TDHF 1}, is simplified as
\begin{equation}
    \sum_{\textbf{k}^\prime}\left[\omega I_{\textbf{k}\textbf{k}^\prime} - \left( E_{\textbf{k}\textbf{k}^\prime} + \sum_{\textbf{p}^\prime} C_{\textbf{k}\textbf{p}^\prime} V_{\textbf{p}^\prime\textbf{k}^\prime} \right) \right]\chi_{\textbf{k}^\prime \textbf{p}}(\omega)=C_{\textbf{k}\textbf{p}}.
\end{equation}
The excitation energy matrix
\begin{equation}
    E_{\textbf{k}\textbf{k}^\prime} = 
    \begin{bmatrix}
         E_{\textbf{k}+\textbf{q} c\, \textbf{k} v, \, \textbf{k}^\prime+\textbf{q} c\, \textbf{k}^\prime v} &\\&
         E_{\textbf{k} v\, \textbf{k}-\textbf{q} c, \, \textbf{k}^\prime v\, \textbf{k}^\prime-\textbf{q} c }  
    \end{bmatrix}=
    \begin{bmatrix}
        \delta_{\textbf{k}\textbf{k}^\prime}(E_{\textbf{k}+\textbf{q}c}-E_{\textbf{k}v}) & \\ & -\delta_{\textbf{k}\textbf{k}^\prime}(E_{\textbf{k}-\textbf{q}c}-E_{\textbf{k}v})
    \end{bmatrix},
\end{equation}
and the commutation matrix
\begin{equation}
    C_{\textbf{k}\textbf{p}} = 
    \begin{bmatrix}
        & C_{\textbf{k}+\textbf{q} c\, \textbf{k} v, \, \textbf{p} v \,\textbf{p}+\textbf{q} c } \\
         C_{\textbf{k} v\, \textbf{k}-\textbf{q} c, \, \textbf{p}-\textbf{q} c \,\textbf{p} v } & 
    \end{bmatrix}=
    \begin{bmatrix}
        & I \\
         -I & 
    \end{bmatrix} = ZX,
\end{equation}
where $Z$ and $X$ are Pauli matrices. 

The interaction matrix $V$ 
\begin{equation}
        V_{\textbf{p}\textbf{k}} = 
    \begin{bmatrix}
        V_{\textbf{p}-\textbf{q} c\, \textbf{p} v, \, \textbf{k}+\textbf{q} c \,\textbf{k} v } & 
        V_{\textbf{p}-\textbf{q} c\, \textbf{p} v, \, \textbf{k} v \,\textbf{k}-\textbf{q} c } \\
         V_{\textbf{p} v\, \textbf{p}+\textbf{q} c, \, \textbf{k}+\textbf{q} c \,\textbf{k} v } & 
         V_{\textbf{p} v\, \textbf{p}+\textbf{q} c, \, \textbf{k} v \,\textbf{k}-\textbf{q} c }
    \end{bmatrix}.
\end{equation}
The momenta of the elements are all different, and we need to calculate them 
in the valley/spin $s$-representation and then transform to the energy eigenstate $n$-representation ($n=c/v$).
For the top-left block,
\begin{equation}
\begin{aligned}
    &V_{\textbf{p}-\textbf{q} s_1\, \textbf{p} s_2, \, \textbf{k}+\textbf{q} s_3 \,\textbf{k} s_4 }\\
    =& \ v_{\textbf{p}-\textbf{q} s_1\ \textbf{k}+\textbf{q} s_3\ \textbf{k} s_4\ \textbf{p} s_2} - v_{\textbf{k}+\textbf{q} s_3\ \textbf{p}-\textbf{q} s_1\ \textbf{k} s_4\ \textbf{p} s_2}\\
    =&\ \ \ \ \delta_{s_1 s_2} \delta_{s_3 s_4} \frac{1}{A} \sum_{\textbf{q}^\prime} V(\textbf{q}^\prime) 
    \left< \textbf{k}+\textbf{q} \middle| e^{+i\textbf{q}^\prime \cdot \textbf{r}} \middle| \textbf{k} \right>_{s_{34}}
    \left< \textbf{p}-\textbf{q} \middle| e^{-i\textbf{q}^\prime \cdot \textbf{r}} \middle| \textbf{p} \right>_{s_{12}}\\
    &-\delta_{s_1 s_4} \delta_{s_3 s_2} \frac{1}{A} \sum_{\textbf{q}^\prime} V(\textbf{q}^\prime) 
    \left< \textbf{k}+\textbf{q} \middle| e^{+i\textbf{q}^\prime \cdot \textbf{r}} \middle| \textbf{p} \right>_{s_{23}}
    \left< \textbf{p}-\textbf{q} \middle| e^{-i\textbf{q}^\prime \cdot \textbf{r}} \middle| \textbf{k} \right>_{s_{14}}\\
    =& \ \ \ \ \delta_{s_1 s_2} \delta_{s_3 s_4} \frac{1}{A} \sum_{\textbf{G}} V(\textbf{q}+\textbf{G}) F^{s_{34}}(\textbf{q}+\textbf{G}) F^{s_{12}}(-\textbf{q}-\textbf{G}) e^{i\big( (s_{12}\textbf{p}-s_{34}\textbf{k} ) \times \textbf{G} \big)_zl_B^2  } e^{\frac{i}{2} \big\{ \textbf{q} \times \big[ s_{12}(-\textbf{p}-\textbf{G}) + s_{34}(\textbf{k}-\textbf{G}) \big] \big\}_z l_B^2} \\
    &-\delta_{s_1 s_4} \delta_{s_3 s_2} \frac{1}{A} \sum_{\textbf{G}} V(\textbf{k}-\textbf{p}+\textbf{q}+\textbf{G}) F^{s_{23}}(\textbf{k}-\textbf{p}+\textbf{q}+\textbf{G}) F^{s_{14}}(-\textbf{k}+\textbf{p}-\textbf{q}-\textbf{G}) \\
    &\ \ \ \ \ \ \ \ \ \ \ \ \ \ \ \ \ \ \ \ \ \ \ \ \ \ \ \ \ \ \ \ \ \ \ \ \ \ \ \ \ \ \ \ \ \ \ \ \ \ \ \ \ \ \ \ \times\ 
    e^{\frac{i}{2} (s_{14}-s_{23}) \big(\textbf{p}\times\textbf{k}+(\textbf{p}+\textbf{k})\times\textbf{G}\big)_z l_B^2 } e^{\frac{i}{2} \big\{ \textbf{q} \times \big[ s_{23}(\textbf{p}-\textbf{G}) + s_{14}(-\textbf{k}-\textbf{G}) \big] \big\}_z l_B^2 },
\end{aligned}
\end{equation}
where $s=\pm1$ and $F(\textbf{q})$ depends on the Landau level that we work on.
For the other three blocks, we need to shift the momenta before the representation transform.
An extra phase in eq.~\ref{eq:boost by G} emerges when the shift crosses the Brillouin zone boundary.

Although $E+CV=E+ZXV$ is not hermitian, $ZE+XV$ is hermitian because $V_{\tau\sigma,\mu\nu} = (V_{\nu\mu,\sigma\tau})^*$ by its definition eq.~\ref{eq: matrix V}. 
Furthermore, one can prove that it is exactly the second-order derivative, the Hessian, of the energy functional with respect to the density matrix.
In the HF ground state, it must be positive (semi-)definite and can be (pivoted) Cholesky decomposed as $ZE+XV = L L^\dagger$, where $L$ is an lower-triangular matrix.
Therefore, 
\begin{equation}
\begin{aligned}
    \Big[ \omega I - Z L L^\dagger \Big]\chi(\omega) & = ZX\\
    L^\dagger \cdot \Big[ \omega I - Z L L^\dagger \Big]\chi(\omega) \cdot L & = L^\dagger ZX L\\
    \Big[ \omega I - L^\dagger  Z L \Big]\Big[L^\dagger \chi(\omega) L\Big] & = L^\dagger ZX L\\
\end{aligned}
\end{equation}
By such a transform, the eigen problem of a non-hermitian matrix $E+CV = Z L L^\dagger$ is converted to one of a Hermitian matrix $L^\dagger  Z L$ with the eigenvalues invariant.

\twocolumngrid
\bibliography{FQSH1.bib}

\end{document}